\documentclass[12pt,a4paper]{article}

  \usepackage{a4wide}
  \usepackage{latexsym}
  \usepackage{epsf}
  \usepackage{amssymb}
  \usepackage{graphicx}
  \usepackage{amsmath, cite}
  \usepackage{slashed,epsfig}
  \usepackage{bbm}
  \usepackage{bbm}
  \usepackage{amsmath,amssymb,amsthm}
  \usepackage{pst-all}
  \usepackage{here}

\renewcommand{\d}{\textrm{d}}
\newcommand{\Real}{\textrm{I\!R}}
\def\rme{{\mathrm e}}
\newcommand{\e}{\textrm{e}}
\renewcommand{\d}{\textrm{d}}

\newcommand{\SO}{\mathop{\rm SO}}

\newcommand{\Sl}{\mathop{\rm {}S}\ell }

\begin{document}

\begin{flushright}
\small
UUITP-19/10\\
UG-10-20\\
\date \\
\normalsize
\end{flushright}

\begin{center}

\vspace{.7cm}

{\LARGE {\bf Black holes in supergravity and integrability}} \\

\vspace{1.2cm}

{\large W. Chemissany$^a$,  P. Fr\'e$^b$, J. Rosseel$^{c}$,\\
A. S. Sorin$^{d}$, M. Trigiante$^{e}$ and  T. Van Riet$^f$} {}~\\
\vspace{1cm}

 $^a$ {\small\slshape University of Lethbridge, Physics Dept., Lethbridge
T1K 3M4, Alberta, Canada }\\\vspace{0.2cm}

$^b$ {\small\slshape Dipartimento di Fisica Teorica, Universit\'a di
Torino, \\
$\&$ INFN - Sezione di Torino\,, via P. Giuria 1, I-10125 Torino,
Italy}\\\vspace{0.2cm}

$^c$ {\small\slshape Centre for Theoretical Physics, University of Groningen,\\
    Nijenborgh 4, 9747 AG Groningen, The Netherlands}\\ \vspace{0.2cm}

$^d${\small\slshape Bogoliubov Laboratory of Theoretical
Physics, Joint Institute }\\
{\small\slshape for Nuclear Research, 141980 Dubna, Moscow Region,
Russia}\\\vspace{0.2cm}

$^e${\small\slshape Dipartimento di Fisica Politecnico di Torino,\\
C.so Duca degli Abruzzi, 24, I-10129 Torino, Italy}\\\vspace{0.2cm}

$^f$ {\small\slshape Institutionen f\"{o}r Fysik och Astronomi, Box 803, SE-751 08 Uppsala, Sweden} \\

\vspace{2cm}

{\bf Abstract} \end{center} {\small Stationary black holes of
massless supergravity theories are described by certain geodesic
curves on the target space that is obtained after dimensional
reduction over time. When the target space is a symmetric coset
space we make use of the group-theoretical structure to prove that
the second order geodesic equations are integrable in the sense of
Liouville, by
explicitly constructing the correct amount of Hamiltonians in involution. 
This implies that the Hamilton--Jacobi formalism can be applied,
which proves that all such black hole solutions, including
non-extremal solutions, possess a description in terms of a (fake)
superpotential. Furthermore, we improve the existing integration
method by the construction of a Lax integration algorithm that
integrates the second order equations in one step instead of the
usual two step procedure. We illustrate this technology with a
specific example
}

\newpage

\pagestyle{plain} \tableofcontents

\section{Introduction}
The study and construction of black hole solutions in supergravity
has a long history \cite{Mohaupt:2000gc, Andrianopoli:2006ub,
Pioline:2008zz, Ferrara:2008hwa, Bellucci:2009qv}. A prominent role
in this history is played by solutions that preserve some
supersymmetry.  Such solutions are easier to construct because the
preservation of some fraction of supersymmetry leads to first order
equations  instead of the usual second order equations of motion.
Furthermore, supersymmetric solutions are protected, to some extent,
from quantum corrections and this allowed to compare supergravity
states with open string states, and thus to verify the celebrated
Bekenstein-Hawking law for the black hole entropy from direct
microstate counting.\par
However, non-supersymmetric solutions are perhaps even more
interesting. The lack of first order BPS equations, that arise from
demanding preservation of supersymmetry, implies that we need to
find other tools to integrate the second order equations of motion
in this case. This paper is concerned with providing such tools. Our
analysis will be restricted to spherically symmetric, asymptotically
flat black holes.

The second order differential equations can be derived from two
different, but equivalent, effective actions and the tools that are
developed for studying non-supersymmetric solutions depend on the
choice of the effective action. Let us briefly summarize these matters,
assuming the simplest case of stationary, spherically symmetric
(single-centered) black holes. We shall be mainly dealing with static solutions in four dimensions,  although our analysis in three dimensions will allow us to extend our analysis by introducing a Taub-NUT charge. Furthermore we consider
four-dimensional solutions,  the extension to any other dimension being
straightforward.

The first kind of effective action is obtained, for static  
 solutions, by expressing the
Maxwell field strengths in terms of the quantized  magnetic and
electric charges $m^I$ and $e_I$ via the respective equations of
motion (and Bianchi identities). Then all remaining degrees of
freedom depend on the radial coordinate $\tau$, rendering the
effective action one-dimensional. The action thus obtained describes
the motion of a particle subject to an external force field
described by the effective, non-positive, potential $-V(\phi,m,e)$
\cite{Gibbons:1996af, Ferrara:1997tw}
\begin{equation}\label{effective1}
\mathcal{L}=\dot{U}^2+ \tfrac{1}{2} G_{rs}\dot{\phi}^r\dot{\phi}^s +
\mathrm{e}^{2\,U}\,V(\phi^r,m,e)\,,
\end{equation}
where $V\ge 0$ is called the effective black hole potential and
$e^{2\,U}$ is the time-component of the static metric. The
$\phi^r$-dependence of $V$ arises from the coupling of the scalars
to the vectors and where we have used $\dot{\phi}^r$ to denote the
derivative of the fields with respect to the radial variable $\tau$.
In this framework the main tool to describe non-supersymmetric
extremal solutions is by mimicking the supersymmetric case. This
means that one tries to find a function $\mathcal{W}_4(U,\,\phi^r;\,m,e)$, named
the ``fake superpotential'' or ``fake central charge'', such that $
2\,\mathrm{e}^{2\,U}\,V=\frac{1}{2}\,(\partial_U\mathcal{W}_4)^2+
\partial_r \mathcal{W}_4\partial^r \mathcal{W}_4$. If this situation is realized the action
can be rewritten as a sum (and difference) of squares. Putting the
separate squares to zero leads to first order flow equations
\begin{equation}
\dot{\phi}^r=G^{rs}\partial_s \mathcal{W}_4\,\,\,,\,\,\,\,\,\dot{U}=\partial_U
\mathcal{W}_4\,.
\end{equation}
This is nothing else but the familiar Hamilton--Jacobi integration
method \cite{Andrianopoli:2009je}, and $\mathcal{W}_4$ is the Hamilton's
characteristic function associated with the autonomous system
(\ref{effective1}). Such road to the understanding of non-susy
extremal solutions has been used in e.g. \cite{Ceresole:2007wx,
LopesCardoso:2007ky, Andrianopoli:2007gt, Ferrara:2009bw,
Bossard:2009we, Ceresole:2009vp, Ceresole:2009iy} and references
therein. It has later been realized, starting with
\cite{Miller:2006ay}, that this approach could be extended to the
non-extremal Reissner-Nordstr\"om black hole, and it has
subsequently been generalized to other, more involved, non-extremal
solutions \cite{Cardoso:2008gm, Janssen:2007rc,
Perz:2008kh}\footnote{ The most general form of the non-extremal
flow equations were derived in \cite{Perz:2008kh}.}.

In this paper we use the second form of the effective action, found
in \cite{Breitenlohner:1987dg}, which is obtained by dimensional
reduction of the four-dimensional action over the timelike direction
to three dimensions. In three dimensions all vectors can be dualised
to scalars such that one ends up with a supergravity theory, whose
bosonic part of the action is given by
\begin{equation} \label{nonlinsigma}
S = \int \, \d^3 x\, \sqrt{g_3}\Bigl( \tfrac{1}{2} R_3 -
\tfrac{1}{2}g_{ij}(\phi)\partial\phi^i\partial\phi^j\Bigr)\,.
\end{equation}
The scalars $\phi^i$ parametrize a target space with metric
$g_{ij}(\phi)$ of indefinite signature
\cite{Cremmer:1998em,Cortes:2003zd,Hull:1998br,Cortes:2009cs}. One
can show that gravity decouples from the sigma model such that the
non-trivial part of the effective action becomes a pure geodesic
problem
\begin{equation}\label{effective2}
\mathcal{L}=\tfrac{1}{2}g_{ij}\dot{\phi}^i\dot{\phi}^j\,.
\end{equation}
The target space in three dimensions (with metric $g_{ij}$) contains
the target space in four dimensions (with metric $G_{rs}$) as a
subspace. This description of supergravity solutions as geodesic
curves goes beyond four-dimensional black holes and works for
generic supergravity solutions that depend effectively on one
direction, like e.g. stationary and cosmological $p$-brane
solutions, wormholes, instantons, and so on (see
\cite{Bergshoeff:2008be} and references therein).

Similar to the first effective action (\ref{effective1}), the
Hamilton--Jacobi formalism, if it can be applied, leads to a (fake)
superpotential $\mathcal{W}_3(\phi^i)$ such that the second order differential
equations can be integrated to first order flow equations
\begin{equation}
\dot{\phi}^i=g^{ij}\partial_j \mathcal{W}_3\,.
\end{equation}
As explained in \cite{Andrianopoli:2009je}, $\mathcal{W}_4(U,\phi^r;\,m,e)$ is related to
$\mathcal{W}_3(\phi^i)$ in a simple manner. We shall mainly be concerned with the latter, which  
we shall simply denote by $\mathcal{W}$.

Both approaches (\ref{effective1})nd (\ref{effective2}) are
equivalent, but the second approach (\ref{effective2}) is beneficial
since it makes more symmetries apparent. Moreover, when the target
space in three dimensions is a symmetric space (as is often the case
in extended supergravities\footnote{The target spaces in
supergravity are symmetric for theories with more than 8
supercharges. An interesting subset of theories with 8 or less
supercharges exhibits symmetric target spaces as well.}), one can
use group theoretical algorithms to integrate the second order
equations of motion, see e.g.\cite{Gal'tsov:1998yu, Gunaydin:2007bg, Breitenlohner:1987dg, Dobiasch:1981vh,
Bergshoeff:2008be,Bossard:2009at, Bossard:2009mz, Bossard:2009bw,
Bossard:2010mv, Gaiotto:2007ag, Fre:2003ep, Fre':2007hd, Fre:2009zz,
Fre:2009dg, Berkooz:2008rj, Chemissany:2007fg, Chemissany:2009af,
Chemissany:2009hq, Fre:2008zd, Fre:2005bs, Gunaydin:2009pk,
Figueras:2009mc, Kim:2010bf,Mohaupt:2010fk}.

These algorithms were first established for the
case that the target space is Riemannian. This case is relevant for
the construction of cosmological solutions in supergravity
\cite{Fre:2003ep, Fre':2007hd, Fre:2008zd, Chemissany:2007fg}. Later
developments have generalized the integration procedures to target
spaces of indefinite signature \cite{Chemissany:2009hq, Fre:2009et,
Chemissany:2009af, Fre:2009dg}, that find applications in describing
stationary solutions such as black holes.
Essentially, these algorithms solve the geodesic equations in a two step process. In
the first step, one relies on the fact that the geodesic equations
can be rewritten as a Lax pair equation, i.e. as a first order
matrix differential equation for the tangent velocities. In the
second step, one eventually solves  the first order system defined
by the expression for the tangent vector.
\par
The fact that the geodesic equations on symmetric spaces can be
written in a Lax pair form, is a strong indication that these
equations are Liouville integrable, since this is a necessary
condition for Liouville integrability \cite{Babelon:1990qk}. In a
recent paper, \cite{Fre:2009et} the Liouville integrability of the
first order equation for the tangent vector was proven, whereas in
this paper we establish the Liouville integrability of the full
second order geodesic equations. This we prove by explicitly
constructing the correct number of conserved quantities (a.k.a
Hamiltonians) that mutually Poisson-commute (a.k.a are in
involution).  These involutive conserved quantities are appropriate
non trivial rational functions of the  Noether charges. The explicit
form of the involutive hamiltonians was discussed in
\cite{Fre:2009et} for the case of $\Sl(N,\mathbb{R})$ algebras (such construction can be shown to work for a variety of other symmetric geometries such as the STU model). Its
general form will be presented in a forthcoming paper
\cite{MARIO.PIETRO.SASHA}. To our knowledge such constructive proof
of Liouville integrability for this class of models is not present
in the existing literature.
\par
Liouville integrability  implies  that there always exists a
description in terms of a (fake) superpotential. This (fake)
superpotential can be lifted to the (fake) superpotential in the
usual four-dimensional formulation that uses a black hole effective
potential \cite{Andrianopoli:2009je}. Therefore, in this
paper we prove that \emph{all} spherically symmetric solutions of symmetric
supergravity theories where the scalar manifold is a homogeneous coset allow a (fake) superpotential\footnote{This
has been conjectured in \cite{Janssen:2007rc}.}. This result is not
only important for the understanding of the differential equations
that govern black hole solutions, but the existence of a fake
superpotential, in the extremal case, comes with specific
implications for the physics of black holes. In particular there are
implications for the asymptotic stability of the attractor points
(for large and small black holes) \cite{Andrianopoli:2010bj}. There
are also implications in the context of holography
\cite{deBoer:1999xf, Hotta:2009bm}.

As suggested by Liouville integrability, we find a novel way to
directly integrate the geodesic equations of motion in a one-step
procedure to find the expression for the coset element. This is a
significant improvement of the existing integration algorithms.

This paper is organised as follows. In section 2 we present the
basic mathematical notions of symmetric spaces and geodesic
equations in Lax pair form. In section 3 the proof of Liouville
integrability of the second order geodesic equations is presented.
Section 4, on the other hand deals with the practical integration
procedure of the geodesic equations. Finally, in section 5 we
illustrate and apply this technology to black hole solutions in
simple supergravity theories. We conclude with a discussion in section 6.

\section{Geodesic curves on symmetric spaces}

\subsection{Symmetric spaces}

Consider a real $n$-dimensional symmetric space $G/H^*$, where $G$
is a semisimple Lie algebra and $H^*$ is a maximal subgroup. For
pseudo-Riemannian manifolds, $H^*$ is a non-compact real form of
some complex semisimple Lie group, whose compact real form will be
denoted by $H$. We denote the Lie algebras of $G$ and $H^*$ by
$\mathfrak{g}$, $\mathfrak{H}^*$ respectively.  The
orthogonal\footnote{For symmetric spaces $G$ is semi-simple and we
define 'orthogonal' with respect to the Cartan--Killing metric of
$\mathfrak{g}$.} complement to $\mathfrak{H}^*$ in $\mathfrak{g}$ is
denoted by $\mathfrak{K}$
\begin{equation} \label{cartandecomp}
\mathfrak{g} =\mathfrak{H}^* \oplus \mathfrak{K} \,.
\end{equation}
This decomposition is called the Cartan decomposition and
alternatively it can be defined by introducing an involutive
automorphism $\theta$, the so-called Cartan involution. This Cartan
involution acts as follows
\begin{equation}
\theta(\mathfrak{H}^*) = \mathfrak{H}^* \,, \qquad \theta(\mathfrak{K})
= - \mathfrak{K} \,.\label{theta}
\end{equation}
As $\theta$ preserves the Lie brackets, we obtain
\begin{equation}\label{KrepH}
[\mathfrak{H}^*,\mathfrak{H}^*] \subset \mathfrak{H}^* \,, \quad
[\mathfrak{K},\mathfrak{K}] \subset \mathfrak{H}^* \,, \quad
[\mathfrak{H}^*,\mathfrak{K}] \subset \mathfrak{K} \,.
\end{equation}
In case one works in a real matrix representation, the action of
$\theta$ can be expressed in terms of a  diagonal matrix $\eta$ (see
e.g. \cite{Bergshoeff:2008be}):
\begin{eqnarray} \label{defeta}
\forall X\in \mathfrak{g}&:&\theta(X) = - \eta X^T \eta \,, \qquad
\eta = \mathrm{diag}(\mathbbm{1}_p, - \mathbbm{1}_q) \,.
\end{eqnarray}
Let us also introduce a different decomposition of $\mathfrak{g}$ that
will be useful throughout this article. This is the so-called
Iwasawa decomposition of $\mathfrak{g}$
\begin{eqnarray} \label{iwasawadecomp}
\mathfrak{g}&=&\mathfrak{H}^*\oplus Solv \,,
\end{eqnarray}
where $Solv$ is a maximal  solvable subalgebra of $\mathfrak{g}$.
The corresponding decomposition at the level of group elements does
not hold globally on $G/H^*$, since not all elements $g$ of $G$ can
be expressed in the form $g=s\cdot h^*$, where $s\in \exp(Solv)$ is
a solvable group element and $h^*\in H^*$. \par We denote the
generators of the solvable algebra by $T_A$ \footnote{Or sometimes
by $T_i$. }, the generators of $\mathfrak{H}^*$ by $H_{\alpha}$ and
the generators of $\mathfrak{K}$ by $K_A$. The following relations
then hold
\begin{eqnarray}\label{KA}
K_A&=&\tfrac{1}{2}(T_A +\eta T^T_A \eta)\,,\\
H_{\alpha}&=&\Bigl\{\text{Non-vanishing combinations }\,\,
\tfrac{1}{2}(T_{A} -\eta T^T_{A} \eta)\Bigr\} \,.
\end{eqnarray}
 Choosing a parametrization on $G/H^*$ amounts to defining a coset
representative $\mathbb{L}(\phi)\in G/H^*$.
  It is useful to
choose this representative in the so-called solvable gauge, meaning
that it is obtained by exponentiating a generic element of the
solvable algebra $Solv$ of (\ref{iwasawadecomp})
\begin{equation}
\mathbb{L}(\phi) = \exp\left(\phi^i \,T_i\right)\,,
\end{equation}
where we have used the index $i$ to label the 3$d$ scalar fields and
consequently also used that index for the solvable generators here.
Since, as previously mentioned, the Iwasawa decomposition of $G$
with respect to $H^*$ holds only locally in $G/H^*$, the solvable
coordinates $\phi^i$  do not provide a set of global coordinates but
only span a patch of the whole manifold. However this is not an
issue for black hole solutions as we explain in section
\ref{ssec:solvpar}.\par For the time being, we restrict ourselves to
the solvable patch of $G/H^*$, which is isometric to the solvable
group manifold  $\exp(Solv)$, the metric being defined below. The
vielbein and connection 1-forms, $\mathrm{d}\phi^i\,V_i{}^A$ and
$\mathrm{d}\phi^i\,\mathcal{W}_i{}^\alpha$ resp., are defined via
the decomposition
\begin{eqnarray}
\mathbb{L}^{-1}\mathrm{d}\mathbb{L}&=&
\mathrm{d}\phi^i\,V_i{}^A\,T_A=\mathrm{d}\phi^i\,V_i{}^A\,K_A+\mathrm{d}\phi^i\,\mathcal{W}_i{}^\alpha\,H_\alpha\,.\label{li1f}
\end{eqnarray}
The metric $g_{AB}$ on the tangent space to the manifold at the origin is proportional to the
restriction of the Cartan-Killing metric on $\mathfrak{g}$ to $\mathfrak{K}$
\begin{eqnarray}
g_{ij}&=&g_{AB}\,V_i{}^A\, V_j{}^B\,\,\,\,,\,\,\,\,\,\,g_{AB}=\alpha\,\tilde{\eta}_{AB}\,\,\,\,,\,\,\,\,\,\,\tilde{\eta}_{AB}\equiv {\rm Tr}(K_A\,K_B)\,,\label{gdef}
\end{eqnarray}
where $V_i{}^A$ is the vielbein matrix.
The normalization factor
$\alpha$ depends on the representation of the generators $K_A$ and
is chosen so as to have the standard normalization for the
kinetic terms of the scalars in the Lagrangian. The algebraic structure of $Solv$ is
encoded in its structure constants $f_{AB}{}^C$ and can be either
described by the Maurer-Cartan equations for the 1-forms $V^A\equiv
\mathrm{d}\phi^i\,V_i{}^A$ or by the commutation relations among the
generators $T_A$
\begin{equation}
\d V^A=-\tfrac{1}{2}f^{\quad A}_{BC}V^B\wedge
V^C\,\,\,\,,\,\,\,\,\,\, [T_A, T_B]=f_{AB}^{\quad C}
T_C\,.\label{MCf}
\end{equation}
In view of the following analysis it is useful to define the adjoint
representation on $Solv$ of the coset representative
$\mathbb{L}(\phi^i)\in \exp(Solv)$ and of each single generator
$T_A$
\begin{equation} \label{LminTL}
\mathbb{L}^{-1}T_A\mathbb{L}=\mathbb{L}_A{}^{B}\,
T_B\,\,\,,\,\,\,\,(T_A)_B{}^C=-f_{AB}{}^C\,.
\end{equation}
Then the first equality in (\ref{li1f})  can be written in the
adjoint representation as follows
\begin{equation}
(\mathbb{L}^{-1})_B{}^{D}\partial_i\mathbb{L}_D{}^{C}=-V_i{}^A\,f_{AB}^{\quad
C}\,,
\end{equation}
where $\partial_i\equiv \frac{\partial}{\partial \phi^i}$.

\subsection{Geodesic Lax equations}
Let us now discuss the general features of geodesics on $G/H^*$ and
their description in terms of a Lax pair equation. Consider a
geodesic on $G/H^*$ defined by $n$-functions $\phi^i(\tau)$ of an
affine parameter $\tau$. In analogy with the Riemannian case
\cite{Fre:2005bs}, it was established in
\cite{Chemissany:2009hq,Fre:2009et,Chemissany:2009af} that one can
rewrite the geodesic equations on a symmetric space of indefinite
signature as a matrix differential equation of the Lax form
\begin{equation} \label{laxeq}
\frac{\d}{\d \tau} L = [L,W] \,.
\end{equation}
The Lax operator $L$ and Lax connection $W$ are defined in terms of the pull-back  on the geodesic of the
left-invariant 1-form on $G/H^*$, which we denote by $\Omega $
\begin{eqnarray}
\Omega &\equiv& \mathbb{L}^{-1} \frac{\d}{\d \tau} \mathbb{L} = \dot{\phi}^i\,\mathbb{L}^{-1} \frac{\partial}{\partial \phi^i} \mathbb{L}=Y^A\,T_A=W + L \,,\label{Cartanforms}\\
L & =&  Y^A\,K_A\,,\quad Y^A=\tilde{\eta}^{AB}\text{Tr}(\Omega\, K_B)\,,
\end{eqnarray}
where $\tilde{\eta}^{AB}$ denotes the inverse of
$\tilde{\eta}_{AB}$. From the above definition it follows that the
components $Y^A$ of the Lax matrix $L$ are the pull-back  on the
geodesic of the vielbein 1-forms
\begin{eqnarray}
Y^A&\equiv& \dot{\phi}^i\, V_i{}^A\,.\label{defy}
\end{eqnarray}
Note that $W$ and $L$ depend on the scalars $\phi$ as well
as on their derivatives $\dot{\phi}$ with respect to $\tau$.
Then the geodesic action can be written as
\begin{equation}
S=\int \d \tau\,\mathcal{L}\equiv \int \d \tau\,
\tfrac{1}{2}\,g_{AB}\,Y^A\,Y^B= \int \d \tau
\,\tfrac{\alpha}{2}\,\mathrm{Tr}(LL) \,,\label{geoact}
\end{equation}
where we have used (\ref{gdef}).
When one works in the solvable gauge, one can choose a
representation for which the Lax connection $W$ is given in terms of
$L$ as follows
\begin{equation}
W = L_{>} - L_{<} \,,
\end{equation}
where $L_{>(<)}$ denotes the upper-triangular (resp.
lower-triangular) part of $L$. We are therefore interested in
solving the following Lax equation
\begin{equation} \label{laxeq2}
\frac{\d}{\d \tau} L  + [L_> - L_< ,L] = 0 \,.
\end{equation}
Algorithmic methods to achieve this have been devised in the
mathematical and physical literature
\cite{Fre:2005bs},\cite{Chemissany:2009hq, Fre:2009et,
Chemissany:2009af, kodama-1995,
kodama2-1995,kodama3-1995,Fre:2009dg}. The integration formulae
developed, allow one to obtain an explicit solution
$L_{\mathrm{sol}}(\tau)$ of (\ref{laxeq2}), obeying the initial
condition $L_{\mathrm{sol}}(\tau = 0) = L_0$. In order to extract
the solutions for the scalar fields, one should still solve the
system of differential equations obtained from
\begin{equation} \label{extrasys}
\sum_A \mathrm{Tr} \left(\mathbb{L}(\phi(\tau))^{-1} \frac{\d}{\d
\tau} \mathbb{L}(\phi(\tau)) K_A \right)\tilde{\eta}^{AB} K_B =
L_{\mathrm{sol}}(\tau) \,.
\end{equation}
It turns out that, thanks to the use of the solvable gauge, this
system of differential equations can be  solved iteratively
\cite{Fre:2003ep}.

The fact that the geodesic equations can be written in a Lax pair form
is therefore a strong hint of the integrability of the black hole
equations of motion, viewed as geodesic equations. We should however
keep in mind that the Lax equations are first order, while the
geodesic equations are second order. Only the first integration step
has been shown to be integrable \cite{Fre:2009et}. The full
integration process of the second order equations involves two steps
and integrability of the first step does not yet establish
integrability of the full system.

In section \ref{integrability}, we  give a proof of Liouville
integrability of the full \emph{second order} geodesic equations. In
section \ref{algos}, we give a novel, more practical, way to obtain
solutions of the geodesic equations.

\subsection{Noether charges}
Consider the following matrix
\begin{equation}
Q=\mathbb{L}(\tau)L(\tau)\mathbb{L}(\tau)^{-1}\,.\label{defQ}
\end{equation}
Using the Lax equation one can infer that $Q$ is a constant of motion: $\frac{\d Q}{\d \tau}=0$.
In fact, $Q$ is the matrix of Noether charges and plays an important
role in establishing Liouville integrability. While $L$ is an element of $\mathfrak{K}$, $Q$
is an element of the whole algebra $\mathfrak{g}$.

There is an alternative method to describe the geodesic action in
terms of the symmetric  coset matrix
\begin{equation}
\mathbb{M}(\phi)=\mathbb{L}(\phi)\, \eta \, \mathbb{L}(\phi)^{T}\,,
\end{equation}
where $\eta$ was introduced in (\ref{defeta}). The geodesic action
then reads
\begin{equation}
S=-\frac{\alpha}{8} \int \d
\tau\,\,\text{Tr}\left(\frac{\d}{\d\tau}\mathbb{M}\frac{\d}{\d\tau}
\mathbb{M}^{-1}\right)\,.
\end{equation}
In this language the solution is immediate (see
e.g.\cite{Bergshoeff:2008be})
\begin{equation}\label{M-integration}
\mathbb{M}(\tau) \equiv \mathbb{M}(\phi^i(\tau))=\mathbb{M}(0)\rme^{2 Q^T\,\tau}\,,
\end{equation}
where $\mathbb{M}(0)$ is the matrix $\mathbb{M}$ computed on the values $\phi^i(0)$ of the scalar fields $\phi^i$ at $\tau=0$.
The geodesic is thus uniquely defined by the matrix $Q$ and the initial point $\phi^i(0)$  on the manifold.
From  definition (\ref{defQ}) it follows that $ \mathbb{M}(0)\,Q^T=Q\,\mathbb{M}(0)$. On the other hand, by definition,
 $\mathbb{M}(\tau)$ should be
symmetric for all $\tau$ and the considered solution implements such property. When the initial condition is chosen at
the origin, $\mathbb{M}(0)=\eta$, we find that $Q \in \frak{K}$
\begin{equation}
Q^T=\eta Q\eta\,.
\end{equation}
For generic initial conditions $Q$ is an element of a subspace of
$\mathfrak{g}$ which is isomorphic to $\frak{K}$. In what follows we
choose the moduli such that $\mathbb{M}(0)=\eta$.

Equation (\ref{M-integration}) is an explicit integration of the
second order equations, however, not a useful one for matrices with
dimension larger than two as the problem of extracting the
individual scalar fields from the solution for $\mathbb{M}$ becomes
non-trivial. Below we present an integration procedure that gives
the explicit solution for the coset representative $\mathbb{L}$.
This can then be used to find the profile of the individual scalars.
The latter procedure is simpler than extracting the scalars out of
$\mathbb{M}$ because of the upper-triangular structure of
$\mathbb{L}$. The integration procedure we present below for
$\mathbb{L}$ is related to the known Lax integration algorithms.
\par
Fortunately, not all questions of black hole physics require the
exact profile of the various scalar fields such that one can learn
already various things in this framework. For instance, it is useful
to understand the space of black hole solutions and its various
subspaces, such as 1) the space of all regular solutions, 2) the
space of supersymmetric solutions, 3) the space of extremal, regular
non-supersymmetric solutions, and so on. It turns out that these
spaces can all be characterized by simple constraints on $Q$ as we
explain in section \ref{applications}.
\par

\section{Integrability} \label{integrability}

\subsection{Preliminary facts}
Consider the geodesic Lagrangian defined in (\ref{geoact})
\begin{equation}
\mathcal{L}=\tfrac{1}{2}\,g_{AB}\,Y^A\,Y^B=\tfrac{1}{2}\,g_{ij}\dot{\phi}^i\dot{\phi}^j\,.
\end{equation}
Note that the above  Lagrangian describes an autonomous system in
which the radial variable $\tau$ plays the role of time. We define
the momenta $P_i$ conjugate to $\phi^i$ and the Hamiltonian
function
\begin{eqnarray} \label{defPH}
P_i&\equiv & \frac{\partial \mathcal{L}}{\partial
\dot{\phi}^i}=g_{ij}\,\dot{\phi}^j\,\,\,,\,\,\,\, \mathcal{H}=
\tfrac{1}{2}\,g^{ij}\,P_i\,P_j=\mathcal{L}\,.
\end{eqnarray}
The variables $\phi^i,\,P_j$ span the phase space of the system and
will be collectively denoted $Z=\{\phi^i, P_j\}$. The Poisson
bracket on phase space is defined as usual
\begin{equation}
\{\phi^i,\phi^j\}=0\,,\qquad \{P_i, P_j\}=0\,\qquad \{P_i, \phi^j\}=
- \delta_i^j\,.\label{canpp}
\end{equation}
We can write the geodesic equations of motion in a compact form as follows
\begin{equation}\label{Hamilton}
\dot{Z} + \{\mathcal{H}, Z \} =0\,.
\end{equation}
From their definition (\ref{defy}) we derive the expression of $Y^A$
 in terms of the conjugate variables
 \begin{equation}
Y^A = g^{AB}\,V_B{}^i\,P_i\,,\label{YP}
 \end{equation}
 where $V_A{}^i$ denotes the inverse vielbein matrix ($V_A{}^i\, V_B{}^j\,g^{AB}=g^{ij}$)
of the solvable group manifold. From (\ref{canpp}) and
(\ref{YP}) and the Maurer-Cartan equations (\ref{MCf}) one verifies the
following Poisson brackets
\begin{equation}\label{canonical}
\{Y_A, Y_B\}=-f_{AB}^{\quad C}
Y_C\,\,\,,\,\,\,\,\{F(\phi^i),\,Y_A\}=V_A{}^i\,\frac{\partial
F}{\partial \phi^i}\,,
\end{equation}
where we have defined  $Y_A\equiv g_{AB}\,Y^B$, which can be viewed
as a basis for the dual solvable Lie algebra $Solv^*$. This  is
the natural Poisson bracket induced by the algebra\footnote{For any
given Lie algebra $\frak{g}$ we can turn the dual Lie algebra
$\frak{g}*$ into a Poisson manifold since there exists a natural way
to define the Poisson bracket. Consider two functions $F_1,F_2$ on
$\frak{g}*$ and take $Y\in \frak{g}*$, then we define the bracket
$\{.,.\}$ as $\{F_1,\,F_2\}(Y)\equiv (\partial^A F_1)(\partial^B
F_2) f_{AB}{}^{ C}\,Y_C$, with $\partial^A F\equiv \frac{\partial
F}{\partial Y_A}$. This Poisson bracket indeed reproduces
(\ref{canonical}) if the representation is taken with the minus sign
$T_A\rightarrow -T_A$. This was used in \cite{Fre:2009dg} to prove
Liouville integrability of the first order problem for the geodesic
tangent velocity vector.}.


%
%
%
Let us now define the following $n$ components of the Noether charge
matrix $Q$
\begin{equation}
Q_A\equiv \alpha \,\textrm{Tr}(Q\,T_A)\,.\label{QA}
\end{equation}
From the definition (\ref{defQ}) of $Q$ one can derive the following
relation between $Q_A$ and $Y_A$
\begin{equation}
Q_A \equiv \mathbb{L}_A{}^{B}\, Y_B\,.\label{QY}
\end{equation}
 Aside from being constants of motion,  the above relation and eq. (\ref{canonical})  imply that
$Q_A$ are in involution with the $Y^A$
\begin{equation}\label{bracket1}
\{Q_A,\,Y_B\}=0\,,
\end{equation}
and that, moreover, they satisfy the following Poisson relations
\begin{equation}
\{Q_A,\,Q_B\}=f_{AB}{}^C\,Q_C\,. \label{bracket2}
\end{equation}
Hence, whereas the $Y_A$ are related to the Killing vectors
associated with the right action of $Solv$, the $Q_A$ are related to
the left action \cite{Olive}.

The $Y_A$ and $Q_A$ introduced above depend on the phase space
variables $Z$ via their definitions. In the next section, we show
that one can construct a number of constants of motion that depend
on $Y_A$ and $Q_A$. These constants Poisson-commute and one can find
the correct number of them, so as to constitute a complete set of
Hamiltonians as required by Liouville integrability.

\subsection{The proof of Liouville integrability}\label{proofIn}
 Liouville integrability, see for instance \cite{Arnold}, is the statement that
there exist $n$ functionally independent constants of motion
$\mathcal{H}_i(Z)$, here referred to as Hamiltonians, that
Poisson-commute with each other
\begin{equation} \label{commC}
\{\mathcal{H}_i, \mathcal{H}_j\}=0\,.
\end{equation}
This statement implies the usual Hamilton--Jacobi formulation of
integrability in terms of a \emph{(fake) superpotential $W$}, which
is the language used in the supergravity literature. This is
explained below in subsection \ref{HJ}.\par
What has been established sofar in \cite{Fre:2009et} is the
Liouville integrability of the \emph{first order} problem
\begin{equation}\label{FIRST}
\dot{Y}_A + \{\mathcal{H}, Y_A\}=0\,.
\end{equation}
Since the Poisson bracket on the dual Lie algebra $Solv^*$  (the
space spanned by the $Y_A$) is degenerate, integrability implies the
existence of a symplectic foliation for which the Hamiltonian flow
is Liouville integrable on the symplectic leaves. Each leaf is
nothing but the co-adjoint orbit of an element $(Y_A)$ of $Solv^*$.
Denoting the
dimension of the coset by $n$ and the dimension of the leaves by
$2h_O$ (since the symplectic leaves are always even dimensional),
we have by definition
\begin{eqnarray}
2h_O&=&{\rm rank}(f_{AB}{}^C\,Y_C)\,.
\end{eqnarray}
Reference \cite{Fre:2009et} showed the existence of $(n -h_O)$
constants of motion in involution, where $h_O$ of them correspond to
Hamiltonians in involution on the symplectic leaf and the remaining
$n-2h_O$ constants are referred to as \emph{Casimirs}.
A Casimir $\mathcal{H}(Y_A)$ is defined by the property
\begin{eqnarray}
\{\mathcal{H}(Y_A),\,Y_B\}&=&0\,\,\,,\,\,\,\,\,\forall \ B=1,\dots, n\,.\label{Casimirr}
\end{eqnarray}
The Casimirs define the foliation since the symplectic leaves are labeled by the
values of a maximal set of functionally independent Casimir functions.
For the moment we are not
interested in the exact form and construction of these Hamiltonians;
we simply assume their existence. The explicit expression for the
Hamiltonians can be found in \cite{Fre:2009et} and we recall the
relevant formulas in section \ref{applications}, when we need them
explicitly.

Let us denote the Hamiltonians on the leaves by $\mathcal{H}_a(Y_A)$,
where $a=1,\ldots,h_O$ and the Casimirs by
$\mathcal{H}_\ell(Y_A)$, where $\ell=1,\ldots, n-2h_O$. If we use the
identities (\ref{bracket1}) and (\ref{bracket2}) we find $2(n
-h_O)$ constants of motion, which Poisson-commute
\begin{equation}
\mathcal{H}_a(Y_A)\,,\quad \mathcal{H}_\ell(Y_A)\,,\quad
\mathcal{H}_a(Q_A)\,,\quad \mathcal{H}_\ell(Q_A)\,,
\end{equation}
where the $\mathcal{H}_a(Q_A)$, resp. $\mathcal{H}_\ell(Q_A)$ are
obtained by replacing $Y_A$ by $Q_A$ in $\mathcal{H}_a(Y_A)$,
$\mathcal{H}_\ell(Y_A)$.
One can show that each  $\mathcal{H}_\ell(Q_A)$ is itself a Casimir, namely it satisfies eq. (\ref{Casimirr})
 and can thus be expressed as a function of the $\mathcal{H}_\ell(Y_A)$.
 To prove this, consider a generic representative  $\mathcal{H}_\ell(Q_A)$ of this set.
  By construction it satisfies the equations obtained by replacing $Y_A\rightarrow Q_A$ in (\ref{Casimirr})
\begin{eqnarray}
\frac{\partial \mathcal{H}_\ell}{\partial
Q_A}(Q_E)\,f_{AB}{}^C\,Q_C=0\,\,\,,\,\,\,\,\,\forall \ B=1,\dots,
n\,.\label{CasimirrQ}
\end{eqnarray}
Using the relation (\ref{QY}) and the invariance of $f_{AB}{}^C$ under the action of $\mathbb{L}\in \exp(Solv)$, we conclude that
$\mathcal{H}_\ell(Q_A)=\mathcal{H}_\ell(\mathbb{L}_A{}^B\,Y_B)$, as a function of $Y_B$, satisfy eq.  (\ref{Casimirr}),
and thus correspond to Casimirs.\par
 The remaining $\mathcal{H}_a(Q_A)$ are independent of the
$\mathcal{H}_a(Y_A)$ and $\mathcal{H}_\ell(Y_A)$. This gives us a total
of
\begin{equation}
(n-h_O) + h_O = n\,,
\end{equation}
Hamiltonians in involution, thereby proving Liouville integrability
of the second order problem. The Hamiltonians $\mathcal{H}_i$ of eq.
(\ref{commC}) are thus explicitly constructed as $\{\mathcal{H}_i\}
= \{\mathcal{H}_a(Y) , \mathcal{H}_\ell(Y), \mathcal{H}_a(Q)\}$.

\subsection{From Liouville to Hamilton--Jacobi and (fake)
superpotentials}\label{HJ}
In order to fill the gap between the classical mathematical language of Liouville integrability and the language adopted in the current supergravity literature on black hole solutions, it is convenient to recall a few basic definitions and concepts concerning the \textit{momentum map} on symplectic manifolds.
\par
Let  $\mathcal{M}_{2n}$ be a real even-dimensional manifold endowed with a closed non-degenerate two-form:
\begin{eqnarray}
    \Omega & = & \Omega_{\alpha\beta} (Z) \, dZ^\alpha \, \wedge \, dZ^\beta \, ,\nonumber\\
    d \Omega & = & 0 \label{dueforma}
\end{eqnarray}
where $Z^\alpha$ denote the $2n$ coordinates in any given  patch.  The pair $\left( \mathcal{M}_{2n}\, , \, \Omega \right)$ is named a symplectic manifold. Consider moreover vector-fields on the manifold $\mathcal{M}_{2n}$
\begin{equation}\label{vecticanonical}
    \mathbf{X} \, = \, X^\alpha \, \frac{\partial}{\partial Z^\alpha}\,,
\end{equation}
with the property that they respect the symplectic structure induced by $\Omega$. This means that the Lie-derivative of $\Omega$ along $\mathbf{X}$ vanishes:
\begin{equation}\label{trullo}
   0 \, = \, \ell_{\mathbf{X}} \Omega\, \equiv \, {\it i}_{\mathbf{X}}d\Omega \, + \, d {\it i}_{\mathbf{X}}\Omega\,.
\end{equation}
In the above equation the symbol ${\it i}_{\mathbf{X}}$ denotes contraction of the given form along the mentioned vector field. Vector fields fulfilling eq.(\ref{trullo}) are named \textit{symplectic}. It follows from this definition that for any symplectic vector field $\mathbf{X}$ the one--form ${\it i}_{\mathbf{X}}\Omega$ is closed and hence locally exact on any open neighborhood  $\mathcal{U} \subset \mathcal{M}_{2n}$. In other words on any  $\mathcal{U}$ we can construct a function $\mathfrak{P}^{(\mathcal{U})}_{\mathbf{X}}$ which solves the following equation:
\begin{equation}\label{momentummap}
    {\it i}_{\mathbf{X}}\Omega|_\mathcal{U} \, = \, d\mathfrak{P}^{(\mathcal{U})}_{\mathbf{X}}|_\mathcal{U}\,.
\end{equation}
The map from the tensor product of the tangent space  $T\mathcal{U}$ with $\mathcal{U}$ into the real numbers:
\begin{equation}\label{momentP}
   \mathfrak{P}^{(\mathcal{U})}_{\mathbf{X}} \, : \, T\mathcal{U} \, \times \, \mathcal{U} \mapsto \, \mathbb{R}
\end{equation}
is named the momentum-map. Clearly the non-trivial topology of the symplectic manifold reflects  itself into the fact that for the same vector field $\mathbf{X}$ the momentum-map on the intersection of different  patches $\mathcal{U}$ and $\mathcal{U}'$ can be related by a non-trivial transition function $f_{\mathcal{U}\mathcal{U}'} \, : \, \mathcal{U}\cap \mathcal{U}' \mapsto \mathbb{R}$. Namely we have:
\begin{equation}\label{transfun}
    \mathfrak{P}^{(\mathcal{U})}_{\mathbf{X}}|_{\mathcal{U}\cap \mathcal{U}'} \, = \, f_{\mathcal{U}\mathcal{U}'} \, \mathfrak{P}^{(\mathcal{U}')}_{\mathbf{X}}|_{\mathcal{U}\cap \mathcal{U}'}.
\end{equation}
Given the existence of the momentum map, one can introduce the Poisson bracket of any two
\textit{functions}\footnote{actually these are not true functions rather they are sections of the line-bundle defined by the transitions functions introduced in eq. (\ref{transfun}). } $\mathfrak{P}_{\mathbf{X}}$ and   $\mathfrak{P}_{\mathbf{Y}}$ associated with two different symplectic vector fields $\mathbf{X}$ and $\mathbf{Y}$\ by means of the following:
\begin{equation}\label{defipois}
    \left \{\mathfrak{P}_{\mathbf{X}} \, , \, \mathfrak{P}_{\mathbf{Y}}\right \} \, \equiv \, \Omega \left( \mathbf{X} \, , \, \mathbf{Y} \right ).
\end{equation}
Note that eq.(\ref{momentummap}) can be interpreted in two ways from left to right or vice-versa. Namely, given a symplectic vector field $\mathbf{X}$ we can construct its momentum-map representation $\mathfrak{P}_\mathbf{X}$ or, given a function $F$ on the manifold $\mathcal{M}$ we can look for the symplectic vector field $\mathbf{X}_F$ such that:
\begin{equation}\label{inversemommap}
    \mathfrak{P}_\mathbf{X_F} \, = \, F
\end{equation}
We can define such a construction  the \textit{inverse momentum map}. \par
A fundamental property of the momentum-map is its \textit{equivariance} which amounts to the following equation:
\begin{equation}\label{equivarianza}
    \forall\,\, \mathbf{X},\mathbf{Y} \, = \, \mbox{symplectic vector fields} \quad : \quad \mathbf{Y} \, \mathfrak{P}_\mathbf{X} \, = \, \mathfrak{P}_{\left [\mathbf{Y},\mathbf{X}\right]}
\end{equation}
Having recalled these concepts let us now return to the case of a canonical system endowed with Liouville integrability. According to our previous definitions this means that on the symplectic manifold $\mathcal{M}$ of dimension $2n$ there exist $n$ independent functions $\mathcal{H}_i(Z)$ in involution, as stated in eq.(\ref{commC}). Let us construct the inverse momentum map of such functions, namely $n$ symplectic vector fields $\mathbf{X}_i$ such that:
\begin{equation}\label{inveliouvil}
    \mathfrak{P}_{\mathbf{X}_i} \, = \, \mathcal{H}_i(Z).
\end{equation}
By construction these vector fields commute with each other and furthermore, in force of equivariance, we have:
\begin{equation}\label{tangent}
    \mathbf{X}_i \, \mathcal{H}_j \, = \, \mathbf{X}_i \mathfrak{P}_{\mathbf{X}_j} \,  = \, \mathfrak{P}_{[\mathbf{X}_i \, , \,\mathbf{X}_j]} \, = \, 0.
\end{equation}
This means that the vector fields $\mathbf{X}_i $ are tangential to the $n$-dimensional level set surface
$\Sigma_{h}$ defined by the following equations:
\begin{equation}\label{livello}
   \mathcal{ H}_i(Z) \, = \, h_i
\end{equation}
where $h_i$ are some set of $n$ real numbers. Since the $n$ independent vector fields $\mathbf{X}_i $ provide a basis of sections of the tangent bundle to the level surface $T\Sigma_{h}$, it follows that the symplectic form $\Omega$ restricted to $\Sigma_{h}$ vanishes. Indeed we have:
\begin{equation}\label{vanishOm}
    \Omega\left(\mathbf{X}_i \,,\, \mathbf{X}_j\right)|_{\Sigma_h} \, = \, \left \{\mathcal{ H}_i \, , \, \mathcal{H}_j \right \} \, = \,0.
\end{equation}
If we use a local  canonical patch  $\mathcal{U}\, \subset \,\mathcal{M}_{2n}$  of  the symplectic manifold in which
\begin{equation}\label{canbasa}
    \Omega \, = \, \sum_i \, dP_i \, \wedge \, d\phi^i \, = \, d( \sum_i \, P_i \, d\phi^i)\,,
\end{equation}
we conclude that on the level surface $\Sigma_h$ we have:
\begin{equation}\label{equazia}
    \left ( d \sum_i \, P_i \, d\phi^i  \right) |_{\Sigma_h}\, = \, 0\,.
\end{equation}
Being closed, the restriction to $\Sigma_h$ of the one-form:
\begin{equation}\label{Pdefi}
    P \, = \, P_i \, d\phi^i\,,
\end{equation}
is  locally exact on the same surface. In other words on each open neighborhood $\mathcal{U}\subset \Sigma_h$ there exists a local function $W^{(\mathcal{U})}$ such that:
\begin{equation}\label{localU}
    P \, = \, dW^{(\mathcal{U})}\,.
\end{equation}
Let us now consider the definition of the level set surface:
\begin{equation}
h_i=\mathcal{H}_i(\phi, P)\,,\label{hP}
\end{equation}
Equations
(\ref{hP}) allow us, formally, to rewrite the momenta as functions
of the $\phi$, provided det$(\partial h_i/\partial P_j)\neq 0$ and use the canonical coordinates $\phi^i$ as
independent coordinates on the level surface $\Sigma_h$. Under these conditions, recalling eq.(\ref{defPH})
we arrive at the conclusions that (locally) we have:
\begin{equation}
\dot{\phi}^i=g^{ij}(\phi)\,P_j(\phi^k,\,h_k)\,,
\end{equation}
which is the more standard Hamilton--Jacobi formulation. The $\mathcal{W}$-function 
$\mathcal{W}(\phi^i,h_j)$ is nothing but the \emph{Hamilton's
characteristic function} solution to the Hamilton-Jacobi equations
\cite{Andrianopoli:2009je,Arnold, McCauley}
\begin{eqnarray}\label{HJW}
\mathcal{H}\left(\frac{\partial \mathcal{W}}{\partial
\phi^i},\phi^i\right)&=&\tfrac{1}{2}\,\frac{\partial \mathcal{W}}{\partial
\phi^i}\,g^{ij}(\phi^k)\,\frac{\partial \mathcal{W}}{\partial \phi^j}=v^2\,.
\end{eqnarray}
We can say that a complete solution $\mathcal{W}(\phi^i,h_j)$ to the above
equation exists in any simply connected domain of the phase space in
which the Jacobian det$(\partial h_i/\partial P_j)\neq 0$ is non
vanishing \footnote{This property is stronger than the so called ``
flow box theorem'', see for instance \cite{Box}, which implies the
existence of a complete solution to the Hamilton-Jacobi equation in
a neighborhood of any point in which $(\frac{\partial
\mathcal{H}}{\partial \phi^i},\,\frac{\partial \mathcal{H}}{\partial
P_i})\neq (0,0)$.}.
\par
In problems related to the study of spherically
symmetric, asymptotically flat black holes, we are not interested in
the complete solution $\mathcal{W}(\phi^i,h_j)$ to the Hamilton--Jacobi
equation, since regularity of the four-dimensional solution implies
severe restrictions on the values of the integrals $h_i$, as we are
going to illustrate in section \ref{reg}.\par
 We shall apply this analysis in section \ref{slthree} to a
specific model and prove that in the domain  spanned by the regular
black hole solutions, the Jacobian det$(\partial h_i/\partial
P_j)\neq 0$ is non-vanishing. The function $\mathcal{W}$ will be explicitly
derived for specific solutions.
\par
\par We would now like to point out
that the integration provided by the Noether charge construction
\begin{equation}
P_i(\phi, Q)=Q_A\,\mathbb{L}^{-1}{}_B{}^{A}\,V_i{}^B\,,
\end{equation}
provides, generically, a non-closed momentum one-form
\begin{equation}
\d P\neq 0\,,
\end{equation}
see \cite{Perz:2008kh} for some examples of this. This means that
the first order integration provided by the Noether charge is
\emph{different} from the first order integration provided by the
Hamilton--Jacobi formalism and therefore the conclusions reached in
\cite{Perz:2008kh} about the existence of the $\mathcal{W}$-function are
incorrect.
\par
In this way we have shown that Liouville integrability  is
completely equivalent to the existence of the so called (fake)
superpotential $\mathcal{W}$. On the other hand, as it was pointed out in
\cite{Fre:2009et} and as it will be further discussed in a next
coming paper \cite{MARIO.PIETRO.SASHA}, all supergravity theories
where the scalar manifold is a generic, not necessarily
symmetric coset manifold, possess Liouville
integrability. Hence for all these cases we proved the existence of
the fake superpotential. Let us also stress that  we can choose one
of the Hamiltonians, say $\mathcal{H}_1$, to coincide with the
quadratic Hamiltonian
$\mathcal{H}(Y^A)=\frac{1}{2}\,Y^A\,Y^B\,g_{AB}$. Its constant value
$h_1$ will also be denoted by $v^2$ in the following and, for black
hole solutions, it represents the extremality parameter.

\section{Lax integration algorithms} \label{algos}

Proving integrability is one thing; constructing the general
integral of the geodesic equations is a separate issue. In this
section, relying on earlier results \cite{Chemissany:2009hq,
Fre:2009et, Chemissany:2009af, kodama-1995, kodama2-1995,
kodama3-1995, Fre:2009dg}, we show that this general integral can be
constructed. All integration algorithms so far developed, focus on
giving a solution for the Lax operator $L(\tau)$. As explained under
(\ref{laxeq2}), this is in principle sufficient, as it allows to
obtain the solutions for the scalar fields after solving a
second system of differential equations that can be solved
explicitly. For practical reasons, it is however desirable to
circumvent this second integration step, by giving an integration
formula for the coset representative. In this section, we will
propose and prove such an integration formula. In section
\ref{laxint1}, we will review how the integration for the Lax
operator is performed. Relying on that result, the formula for the
coset representative will be discussed in section \ref{laxint2}.

\subsection{Integration formula for the Lax operator} \label{laxint1}

Let us repeat here the results of \cite{Chemissany:2009af}. We will
assume that we work in a matrix representation of $\mathfrak{g}$
consisting of $N \times N$-matrices. The authors of
\cite{Chemissany:2009af} showed that the solution for the matrix
elements $L_{pq}(\tau)$ of the Lax operator can be given in closed
form, according to a very specific integration formula that depends
on an initial value $L_0 = L(\tau = 0)$ for the Lax operator. In
fact, several equivalent versions of these formulas were given.

The integration formula for $L(\tau)$ that we start from is given by
\begin{equation}\label{algoronegeneral1}
    L(\tau) \, = \,
   \mathcal{Q}(\mathcal{C})\, L_0\,
   \left(\mathcal{Q}(\mathcal{C})\right)^{-1}\,,
\end{equation}
where the matrix $\mathcal{Q}(\mathcal{C})$ has the following
elements
\begin{eqnarray} \label{Gensolut2}
\mathcal{Q}_{ij}(\mathcal{C}) & \equiv
&\frac{1}{\sqrt{\mathfrak{D}_i(\mathcal{C})\mathfrak{D}_{i-1}(\mathcal{C})}}
\, \mathrm{Det} \, \left(\begin{array}{cccc}
\mathcal{C}_{1,1}(\tau)&\dots &\mathcal{C}_{1,i-1}(\tau)& (\mathcal{C}^{\frac{1}{2}}(\tau))_{1,j}\\
\vdots&\vdots&\vdots&\vdots\\
\mathcal{C}_{i,1}(\tau)&\dots &
\mathcal{C}_{i,i-1}(\tau)& (\mathcal{C}^{\frac{1}{2}}(\tau))_{i,j}\\
\end{array}\right)\,.
\end{eqnarray}
and we have defined
  \begin{equation}\label{cijN}
    \mathcal{C}(\tau)\, := \rme^{-2\, \tau \, L_0}\,,
\end{equation}
\begin{equation}\label{DktN}
    \mathfrak{D}_{i}(\mathcal{C}) \, := \, \mbox{Det} \, \left ( \begin{array}{ccc}
    \mathcal{C}_{1,1}(\tau) & \dots & \mathcal{C}_{1,i}(\tau)\\
    \vdots & \vdots & \vdots \\
    \mathcal{C}_{i,1}(\tau) & \dots & \mathcal{C}_{i,i}(\tau)
    \end{array}\right)  \, , \quad
    \mathfrak{D}_{0}(\tau):=1 \,.
    \end{equation}

The formula (\ref{algoronegeneral1}) can be proven by showing that
it obeys the Lax equation. In order to do this, we start by using
$\mathcal{Q}(\mathcal{C})$ to define the following triangular
matrices
 \begin{eqnarray}
\label{uppertriangl11}
 \mathcal{X}_{>}(\mathcal{C})  \, &:=& \,
  \mathcal{Q}(\mathcal{C})\,\mathcal{C}^{\frac{1}{2}}(\tau)\,, \\
\left(\mathcal{X}_{<}(\mathcal{C})\right)^{-1} \, &:=& \,
  \mathcal{Q}(\mathcal{C})\,(\mathcal{C}^{\frac{1}{2}}(\tau))^{-1}
\,. \label{uppertriangl12}
\end{eqnarray}
One can then show that $\mathcal{X}_>(\mathcal{C})$ (as well as its
inverse) is upper-triangular. Similarly, it can be seen that
$\mathcal{X}_<(\mathcal{C})$  (as well as its inverse) is
lower-triangular. Another useful property of these matrices is that
their diagonal elements are equal
\begin{equation}
(\mathcal{X}_{>}(\mathcal{C}))_{ii} \,=\,
(\mathcal{X}_{<}(\mathcal{C}))_{ii}\,.
 \label{uppertriangl13}
\end{equation}
The proof of (\ref{algoronegeneral1}) then starts by writing the
matrix $\rme^{-\,\tau\,L_0}$ in terms of
$\mathcal{X}_{>}(\mathcal{C})$ and  $\mathcal{X}_{<}(\mathcal{C})$
\begin{eqnarray}
 && \rme^{-\,\tau\,L_0}\,=\,
\left(\mathcal{Q}(\mathcal{C})\right)^{-1}\,\mathcal{X}_{>}(\mathcal{C})\,,
 \label{Cmatrixrepr}\, \\
 && \rme^{-\,\tau\,L_0}\,=\,
 \mathcal{X}_{<}(\mathcal{C})\,\mathcal{Q}(\mathcal{C})\,.
 \label{Cmatrixrepr1}
 \end{eqnarray}
By deriving (\ref{Cmatrixrepr}) and (\ref{Cmatrixrepr1}) with
respect to $\tau$, one can obtain
\begin{eqnarray}
\label{rel1}
 &&\mathcal{Q}(\mathcal{C})\,\frac{d}{d\tau}\,\left(\mathcal{Q}(\mathcal{C})\right)^{-1}\,
 =\, -\,L(\tau)\,-\,\left(\frac{d}{d\tau}\,\mathcal{X}_{>}(\mathcal{C})\right)\,
 \left(\mathcal{X}_{>}(\mathcal{C})\right)^{-1}\, ,\\
&&\mathcal{Q}(\mathcal{C})\,\frac{d}{d\tau}\,\left(\mathcal{Q}(\mathcal{C})\right)^{-1}\,
 =\,+\,L(\tau)\,+\, \left(\mathcal{X}_{<}(\mathcal{C})\right)^{-1}\,
 \left(\frac{d}{d\tau}\,\mathcal{X}_{<}(\mathcal{C})\right)\,. \label{rel2}
 \end{eqnarray}
Using these equations, the triangularity properties of
$\mathcal{X}_>(\mathcal{C})$ and $\mathcal{X}_<(\mathcal{C})$ and
(\ref{uppertriangl13}), one can see that
\begin{eqnarray}
\label{great}
 &&\mathcal{Q}(\mathcal{C})\,\frac{d}{d\tau}\,\left(\mathcal{Q}(\mathcal{C})\right)^{-1}\,
 =\, L_>(\tau)\,-\,L_<(\tau) = W(\tau)\,.
 \end{eqnarray}
By deriving (\ref{algoronegeneral1}) with respect to $\tau$, one
finds
 \begin{equation}
 \frac{d}{d\tau} L(\tau)+\left
[\,\mathcal{Q}(\mathcal{C})\,\frac{d}{d\tau}\,\left(\mathcal{Q}(\mathcal{C})\right)^{-1}\,
, \, L(\tau)\,\right] \, = \, 0\,,
 \label{rel3}
 \end{equation}
so that with (\ref{great}) we find that $L(\tau)$ indeed obeys the
Lax equation.

The formula (\ref{algoronegeneral1}) gives an integration formula
for the Lax operator $L(\tau)$. As explained above, in order to
obtain the explicit expressions for the scalar fields, one still
needs to perform an extra integration. We can actually do better and
give an integration formula for the coset representative
$\mathbb{L}(\tau)$. As the latter is solely expressed in terms of
the scalar fields and no longer in terms of their derivatives, one
can use it to easily extract the scalar field solutions.

\subsection{An integration formula for the coset representative} \label{laxint2}

We now derive an integration formula for the inverse coset
representative. Throughout this derivation, we will assume that we
are working on the coset space $\frac{\Sl(p+q)}{\SO(p,q)}$. This is
not really a restriction, as one can always find values for $p$ and
$q$ and a representation of $G$, such that the coset space $G/H^*$
can be embedded in $\frac{SL(p+q)}{SO(p,q)}$
\begin{equation} \label{slembed}
\frac{G}{H^*} \quad \hookrightarrow \quad \frac{SL(p+q)}{SO(p,q)}
\,.
\end{equation}
As according to this embedding the Lax connection is an element of
the algebra of $\SO(p,q)$, we have
\begin{equation} \label{etaantisym}
W(\tau) \,\eta(p\,,\,q) \, = \, -\,\eta(p\,,\,q)\, W^T(\tau)\,,
\end{equation}
where
\begin{equation}
\eta(p,q)\,:=\,\left(
\begin{array}{ll}
 +{\mathbbm{1}_p} & ~~~0\, \\
 ~~0 & -\,{\mathbbm{1}_q}
\end{array}
\right) \, , \quad p\,+\,q\,=\,N\,.
 \label{etametric}
\end{equation}
Similarly, the Lax operator $L(\tau)$ obeys
\begin{equation}\label{etasym}
 L(\tau) \,\eta(p,q) \, = \, \eta(p,q)\, L^T(\tau) \,.
\end{equation}
We can now use the fact that $L$ and $W$ can be represented
according to the formulae
\begin{eqnarray}\label{L}
L(\tau)\,& = & \,L_>(\tau)\,+\,L_<(\tau)\, -\,\text{diag}(L(\tau))\, \\
W(\tau)\, & = & \, L_>(\tau) - L_<(\tau) \,. \label{W}
\end{eqnarray}
and substituting eqs. (\ref{L}--\ref{W}) into eq.
(\ref{Cartanforms}) we obtain the following relation
\begin{equation}\label{CartanformsSolv}
 \mathbb{L}(\tau)^{-1}\,\frac{d}{d\,\tau}\,\left(\mathbb{L}(\tau)\right) \, = \,2\, L_>(\tau)\,-
 \,\text{diag}(L(\tau))\,.
\end{equation}
On the other hand, let us note that by summing (subtracting) eqs.
(\ref{rel1}) and (from) (\ref{rel2}) one obtains
\begin{eqnarray}
\label{Interest}
 \mathcal{Q}(\mathcal{C})\,\frac{d}{d\tau}\,\mathcal{Q}^{-1}(\mathcal{C})\,
& =&\,
\frac{1}{2}\left[\left(\mathcal{X}_{<}(\mathcal{C})\right)^{-1}\,
 \left(\frac{d}{d\tau}\,\mathcal{X}_{<}(\mathcal{C})\right)\,-\,
 \left(\frac{d}{d\tau}\,\mathcal{X}_{>}(\mathcal{C})\right)\,
 \left(\mathcal{X}_{>}(\mathcal{C})\right)^{-1}\right]\,,\\
L(\tau)\,
 &=&\, -\,\frac{1}{2}\left[\left(\mathcal{X}_{<}(\mathcal{C})\right)^{-1}\,
 \left(\frac{d}{d\tau}\,\mathcal{X}_{<}(\mathcal{C})\right)\,+\,
 \left(\frac{d}{d\tau}\,\mathcal{X}_{>}(\mathcal{C})\right)\,
 \left(\mathcal{X}_{>}(\mathcal{C})\right)^{-1}\right]\,. ~~~~~~~
\label{Interest3} \end{eqnarray} Comparing these equations with
(\ref{great}) allows us to express the right hand side of
(\ref{CartanformsSolv}) as
\begin{eqnarray} \label{Interest1}
&& 2\,L_>(\tau)\,-\,\,\text{diag}(L(\tau))\,=\,
\,\,\mathcal{X}_{>}(\mathcal{C})\, \frac{d}{d\tau}
\,\left(\mathcal{X}_{>}(\mathcal{C})\right)^{-1}\,.
\end{eqnarray}
A simple comparison of equations  (\ref{Interest1}) and
(\ref{CartanformsSolv}) shows that
\begin{eqnarray}\label{solutioncosetrepr}
\left(\mathbb{L}(\tau)^{-1}\right)_{ij}&=&
\left(\mathcal{X}_{>}(\mathcal{C})\,\mathbb{L}(0)^{-1}\right)_{ij}
  \,\\
 &\equiv&\frac{1}{\sqrt{\mathfrak{D}_i(\mathcal{C})\mathfrak{D}_{i-1}(\mathcal{C})}}\mathrm{Det}\left(\begin{array}{cccc} \mathcal{C}_{1,1}(\tau)&\dots
&\mathcal{C}_{1,i-1}(\tau)&
(\mathcal{C}(\tau)\mathbb{L}(0)^{-1})_{1,j}\\
\vdots&\vdots&\vdots&\vdots\\
\mathcal{C}_{i,1}(\tau)&\dots &
\mathcal{C}_{i,i-1}(\tau)& (\mathcal{C}(\tau)\mathbb{L}(0)^{-1})_{i,j}\\
\end{array}\right)\,,\nonumber
\end{eqnarray}
where we have used that from its definition
\begin{eqnarray}\label{X}
\left(\mathcal{X}_{>}(\mathcal{C})\right)_{ij}
  \,\equiv\,\frac{1}{\sqrt{\mathfrak{D}_i(\mathcal{C})\mathfrak{D}_{i-1}(\mathcal{C})}}
\, \mathrm{Det} \, \left(\begin{array}{cccc}
\mathcal{C}_{1,1}(\tau)&\dots &\mathcal{C}_{1,i-1}(\tau)&
\mathcal{C}_{1,j}(\tau)\\
\vdots&\vdots&\vdots&\vdots\\
\mathcal{C}_{i,1}(\tau)&\dots &
\mathcal{C}_{i,i-1}(\tau)& \mathcal{C}_{i,j}(\tau)\\
\end{array}\right).
\end{eqnarray}
Note that we have introduced a constant matrix $\mathbb{L}(0)\,\in
\, \frac{\Sl(p+q)}{\SO(p,q)}$ in the solvable parameterization of
the coset. This matrix can be seen as a matrix of integration
constants. The formula (\ref{solutioncosetrepr}) describes an
explicit general solution for the coset representative
$\mathbb{L}(\tau)\,\in \, \frac{\Sl(p+q)}{\SO(p,q)}$. This solution
is parameterized by the initial data $L_0$ and $\mathbb{L}(0)$,
which gives precisely the correct number of integration constants
associated to the geodesic equations.

\section{Applications to black holes}\label{applications}
\subsection{Black holes as geodesics}
As explained in the introduction, stationary black holes in $d+1$
dimensions can be reduced over time to Euclidean solutions in $d$
dimensions. From now on, we assume that $d=3$, but the extension to
any dimension is straightforward, see e.g. \cite{Bergshoeff:2008be}.

Consider a four-dimensional supergravity describing scalar fields $\phi^r$, and $n_V$ vector fields $B^I$, $I=1,\dots, n_V$.
 We suppose the  scalar manifold  to be homogeneous-symmetric of the form $G_4/H_4$,
 $G_4$ being the isometry group of the manifold and $H_4$ its  maximal compact subgroup.
The action in four dimensions is given by\footnote{We use the more
compact form notation here. A $p$-form $A_p$ is written in components
as $A_p=\tfrac{1}{p!}A_{\mu_1\ldots\mu_p}\d
x^{\mu_1}\wedge\ldots\wedge\d x^{\mu_p}$. The hodge star in $d$
dimensions is defined via, $\star \d x^{\mu_1}\wedge\ldots\wedge\d
x^{\mu_p}=\tfrac{1}{(p-d)!}\epsilon_{\nu_1\,\ldots,\nu_{d-p}}^{\qquad\qquad
\mu_1\ldots\mu_p}\d x^{\nu_1}\wedge\ldots\wedge\d x^{\nu_{d-p}}
$\,,where $\epsilon$ is the totally anti symmetric tensor (not
symbol). This implies that $\star\star A_p=(-1)^{p(d-p)+t}A_p$,
where $t$ is the number of timelike dimensions in the
$d$-dimensional space. We furthermore have the relations $\star
A_p\wedge B_p=\star B_p\wedge
A_p=\tfrac{1}{p!}A_{\mu_1\ldots\mu_p}B^{\mu_1\ldots\mu_p}\star 1$.
The exterior derivative acts as $\d
A_p=\tfrac{1}{p!}\partial_{\mu_0}A_{\mu_1\ldots\mu_p}\d
x^{\mu_0}\wedge\d x^{\mu_1}\wedge\ldots\wedge\d x^{\mu_p}$.}
\begin{equation}\label{4Daction}
S_4=\int\Bigl(\tfrac{1}{2}\star R_4 - \tfrac{1}{2}G_{rs}\star\d
\phi^r\wedge\d\phi^s-\tfrac{1}{2}\mu_{IJ}\star G^I\wedge
G^J+\tfrac{1}{2}\nu_{IJ}G^I\wedge G^J\Bigr)\,,
\end{equation}
where $G^I=\d B^I$, and $G_{rs}$, $\mu_{IJ}$, $\nu_{IJ}$ are
symmetric matrices that depend on the scalars $\phi$; in particular,
$G$ and $\mu$ are required to be positive definite.
  The group $G_4$ represents the on-shell global
symmetry group of the theory, once its non-linear action on the
scalars $\phi^r$ is supplemented by a linear (symplectic)
electric-magnetic duality transformation on the vector field
strengths and their magnetic duals.\par

Let us now restrict ourselves to stationary solutions having an
${\rm SO}(3)$ spatial isometry, which comprise static black holes
though we may allow also for a NUT charge. We then ``reduce'' the
action over the time-like direction using the following ansatz
\begin{align}
& \d s_4^2 =-\e^{2U}(\d t + A_{KK})^2 + \e^{-2U}\d s_3^2\,,\nonumber\\
& B^I= \tilde{B}^I + Z^I(\d t + A_{KK})\,,\label{reductionAnsatz}
\end{align}
where $\tilde{B}^I$ and $A_{KK}$ are vectors in $d=3+0$ and $U$ and
$Z^I$ are scalar fields. Since time is not periodic this is not a
compactification over time, but the difference between a dimensional
reduction over a compact or a non-compact direction only enters the
KK sector, which is assumed truncated here.

When $\d A_{KK}\neq 0$ the 3$d$ vector $A_{KK}$ can not be removed
using a coordinate transformation and this off-diagonal term in the
metric then corresponds to Taub-NUT charge.\par After dimensional
reduction we dualize the vector fields $\tilde{B}^I$ and $A_{KK}$ to
scalars $Z_I,\,a$, see Appendix \ref{APPENDIX} for the details of
this derivation, so to end up with an Euclidean three-dimensional
theory consisting of a sigma model describing $n$ scalar fields
$\phi^i$ coupled to gravity. The fields $\phi^i$ consist of the four-dimensional scalars $\phi^r$, two scalars $U,\,a$ coming from the
four-dimensional metric, and $2\,n_V$ scalar fields $Z^I,\,Z_I$
originating from the four-dimensional vector fields and which will
be collected in a $2\,n_V$-symplectic vector ${\bf Z}\equiv(Z^M)=
(Z^I,\,Z_I)$. The target space of the sigma model is a
homogeneous-symmetric pseudo-Riemannian space of the form $G/H^*$,
$G$ being its isometry group and $H^*$ a non-compact maximal
subgroup of $G$. $G$ is the global symmetry group of the
three-dimensional theory and contains a subgroup of the form ${\rm
SL}(2,\mathbb{R})_E\times G_4$, where ${\rm SL}(2,\mathbb{R})_E$ is
the \emph{Ehlers} group acting transitively on the scalars $U,\,a$.

The ansatz for the solutions in three dimensions is such that the
equations of motion for the scalar fields $\phi^i$, decouple from
the equations for the metric degrees of freedom. To make this clear
we use the following ansatz
\begin{equation}\label{metric1}
\mathrm{d}s^2_3=g_{(3)\,mn}\d x^m\,\d x^n=\exp[4A(\tau)]\mathrm{d}\tau^2
+ \exp[2A(\tau)]\mathrm{d}\Omega_2^2\,,\qquad \phi^i=\phi^i(\tau)\,,
\end{equation}
where $(x^m)=(\tau,\theta,\varphi)$ are the space coordinates and
$\d\Omega_2^2$ is the metric on the unit 2-sphere. The radial
variable $\tau$ is chosen so that
$\sqrt{|g_{(3)}|}\,g_{(3)}^{\tau\tau}$ be $\tau$-independent. Consistently with our
symmetry requirements all the scalar fields will depend on $\tau$
only: $\phi^i=\phi^i(\tau)$. The scalar field equations of motion
can then be derived from the geodesic action (we refer the reader to
Appendix \ref{APPENDIX} for a derivation of this action)
\begin{equation} \label{geodaction}
\mathcal{L} = \tfrac{1}{2}
g_{ij}(\phi)\dot{\phi}^i\dot{\phi}^j=\dot{U}^2+\tfrac{1}{2}\,G_{rs}\,\dot{\phi}^r\,\dot{\phi}^s
+ \tfrac{1}{4}\e^{-4\,U}\,(\dot{a}+{\bf Z}^T\mathbb{C}\dot{{\bf
Z}})^2- \tfrac{1}{2}\e^{-2\,U}\,\dot{{\bf
Z}}^T\,\mathcal{M}_4\,\dot{{\bf Z}}\,,
\end{equation}
where the negative definite, symmetric, symplectic matrix
$\mathcal{M}_4$ is defined in Appendix \ref{APPENDIX}. The target
space metric is indefinite, the negative-signature directions
corresponding to $\d Z^M$.\par
 One thus finds
that $\tau$ corresponds to the affine parameter along the geodesic
and it is a re-parametrization of the usual radial coordinate used
for spherically symmetric solutions. A geodesic that is parametrized
by an affine parameter $\tau$ moves at constant speed, denoted $v^2$
\begin{equation}\label{constant}
\tfrac{1}{2}g_{ij}(\phi)\dot{\phi}^i\dot{\phi}^j\equiv v^2\,.
\end{equation}
Note that $v^2$ can be zero, positive or negative, due to the
indefinite signature of the target space. Note that the effective
Lagrangian (\ref{geodaction}) describes an autonomous system in
which the radial variable $\tau$ plays the role of time and eq.
(\ref{constant}) represents the Hamiltonian constraint. We do not
consider timelike geodesics ($v^2<0$) as they can be shown to
correspond to over-extremal (singular) black hole solutions. Not
every geodesic with $v^2 \geq 0$ corresponds to a regular black hole
solution as we explain in some detail below. For black hole
solutions $v$ is the \emph{extremality parameter} and is expressed
as the product of the black hole temperature $T$ and entropy
$S$\cite{Ferrara:2008hwa}
\begin{equation}
v=2 ST\,.
\end{equation}
Hence, extremal solutions, for which $v=0$, are described by
lightlike geodesics and non-extremal solutions by spacelike
geodesics.

The solution for the metric is independent of the details of the
sigma model and is determined by the geodesic velocity
only\footnote{This expression is sensitive to numerical factors that
depend on the normalisation of the Einstein--Hilbert term with
respect to the normalisation of the scalar kinetic term.}. It can be
found to be
\begin{eqnarray}
 v^2>0 :\qquad
\exp[2A]&=&\frac{v^{2}}{\sinh^2(v\tau)}\,,\label{metric2a}\\
 v^2=0 :\qquad \exp[2A]&=&\frac{1}{\tau^2} \label{metric2b}\,.
\end{eqnarray}
The non-trivial part in finding the three-dimensional solutions
using the ansatz (\ref{metric1}) thus consists of solving the
geodesic equations for the scalar fields $\phi^i$. One of the
scalars in three dimensions is the Kaluza--Klein dilaton $U$ that
comes from the dimensional reduction over time. The
three-dimensional solutions are uplifted to four dimensions using
equations (\ref{reductionAnsatz}).

\subsection{The structure of $\mathfrak{g}$}
For homogenous sigma models that appear upon time-like dimensional
reduction from four dimensions \cite{Hull:1998br,
Bergshoeff:2008be}, the adjoint representation of the isometry group
$G$ branches with respect to its ${\rm SL}(2,\mathbb{R})_E\times
G_4$ subgroup, upon dualization of all the vector fields to scalar
fields as follows
\begin{eqnarray}
{\bf Adj}[G]&\rightarrow &({\bf Adj}[{\rm SL}(2,\mathbb{R})_E],{\bf 1})\oplus ({\bf 1},{\bf Adj}[G_4])\oplus {\bf (2,R)}\,,
\end{eqnarray}
where ${\bf R}$ is the symplectic representation in which the
electric and magnetic charges of the four-dimensional theory
transform  under the duality action of $G_4$.\par The solvable
algebra $Solv$ of (\ref{iwasawadecomp}) has a specific structure
\cite{Bergshoeff:2008be, Fre:2008zd}
\begin{equation}
Solv= Solv_2\oplus Solv_4\oplus
\mathrm{Span}(T_M)\,,\label{solvdec}
 \end{equation}
where $Solv_2=\mathrm{Span}(H_0,\,E_{\beta_0})$ generates a
submanifold ${\Sl}(2,\mathbb{R})/{\SO}(2)$, $\beta_0$ being the
highest root of $\mathfrak{g}$, while $Solv_4$ generates the scalar
manifold $G_4/H_4$ of the four-dimensional parent theory. The
nilpotent generators $T_M$, $M=1,\dots, 2n_V$ correspond to the
positive (restricted) roots $\gamma_M$ defined by the property
$\gamma_M(H_0)=1$. The space $\mathrm{Span}(T_M)$ transforms in the
representation ${\bf R}_{+1}$ with respect to the subgroup
$G_4\times {\rm O}(1,1)$ of $G$, the ${\rm O}(1,1)$ factor being
generated by $H_0$. The algebraic structure of $Solv$ is described
as follows
 \begin{align}
&[Solv_2,\,Solv_4]=0\,\,\,,\,\,\,\,[H_0,\,E_{\beta_0}]=2\,E_{\beta_0}\,,\nonumber\\
&[H_0,\,T_{M}] =
T_{M}\,\,\,,\,\,\,\,[T_M,\,T_N]=\mathbb{C}_{MN}\,E_{\beta_0}\,,\label{algebra}
\end{align}
where $\mathbb{C}_{MN}$ is the symplectic invariant matrix.

At this point we have partially fixed normalisations through
equation (\ref{algebra}). If we furthermore fix the normalisation of
$\mathbb{C}$ such that it has only $\pm 1$ as entries then there is
one scaling freedom left; we can rescale $E_{\beta_0}$ if we
accordingly rescale the $T_M$ generators.

Let us now describe the structure of $\mathfrak{g}$ and of the
subspaces $\mathfrak{K},\,\mathfrak{H}^*$ in the Cartan decomposition
(\ref{cartandecomp}). The matrix $\eta$ defining the decomposition
through the involution $\theta$ (\ref{theta}) has the following
intrinsic expression in terms of the Cartan generator $H_0$:
$\theta=(-1)^{H_0}$. It is easy to verify from (\ref{algebra}) that
\begin{eqnarray}
\eta T_{M} \eta&=&-T_M\,\,\,,\,\,\,\,\,\eta T_{r} \eta=T_r\,\,\,,\,\,\,\,\,\eta E_{\beta_0} \eta=E_{\beta_0}\,.
\end{eqnarray}
For notational convenience we define the basis of generators
$\{T_A\}$ of $Solv$ as follows: $\{T_A\}=\{T_0,T_r,T_M,T_\bullet\}$
where $T_0=\frac{H_0}{2},\,T_\bullet=E_{\beta_0}$ and $\{T_r\}$ is a
basis of $Solv_4$. From the above
properties it follows that the generators $K_A$ of the space
$\mathfrak{K}$, defined in terms of $T_A$ by eq. (\ref{KA}) have the
following form
\begin{eqnarray}
\{K_A\}&=&\{K_0,K_r,K_M,K_\bullet\}\,\,,\,\,\,\,\,K_0=T_0\,\,,\,\,\,K_r=\frac{1}{2}(T_r+T_r^T)\,\,,\,\,\,K_\bullet=
\frac{1}{2}(T_\bullet+T_\bullet^T)\,,\nonumber\\K_M&=&
\frac{1}{2}(T_M-T_M^T)\,.
\end{eqnarray}
The non-compact generators $K_r$ generate the space $\mathfrak{K}_4$
in the Cartan decomposition of the algebra $\mathfrak{g}_4$ of $G_4$
with respect to its maximal compact subalgebra $\mathfrak{H}_4$:
$\mathfrak{g}_4=\mathfrak{H}_4\oplus \mathfrak{K}_4$. The
non-compact generators $\{K_0,\,K_\bullet\}$ generate the space
$\mathfrak{K}_2$ of the Cartan decomposition of the Ehlers algebra
$\mathfrak{sl}(2,\mathbb{R})_E$ with respect to its maximal compact
subalgebra $\mathfrak{u}(1)_E$:
$\mathfrak{sl}(2,\mathbb{R})_E=\mathfrak{u}(1)_E\oplus
\mathfrak{K}_2$. Finally the compact generators $K_M$ span a
$2n_V$-dimensional space $\mathfrak{K}_{{\bf R}_c}$.

The metric $g_{AB}$ at the origin of the manifold is proportional to
the restriction to $\mathfrak{K}$ of the Cartan-Killing metric on
$\mathfrak{g}$, and thus it has $2n_V$  negative signature
directions corresponding to the compact generators $K_M$. The space
$\mathfrak{K}$ transforms, according to equation (\ref{KrepH}) in a
linear representation with respect to the adjoint action of $H^*$. In
particular we can consider the adjoint action on  $\mathfrak{K}$ of
the maximal compact subgroup $H_c$ of $H^*$. This group has the
general form $H_c={\rm U}(1)_E\times H_4$, $H_4$ being the maximal
compact subgroup of $G_4$. The compact symmetry group $H_4$, which
acts linearly on the central and matter charges in four dimensions,
is enhanced in the three-dimensional theory by the ${\rm U}(1)_E$
subgroup of the Ehlers group. With respect to the adjoint action of
$H_c$, the subspace $\mathfrak{K}_{{\bf R}_c}$ spanned by $\{K_M\}$,
transforms in a representation ${\bf R}_c$. This is the
representation in which the central and matter charges in four
dimensions transform with respect to $H_c$. For instance, in the
maximally supersymmetric theory, $G_4={\rm E}_{7(7)}$, $H_4={\rm
SU}(8)$, $G={\rm E}_{8(8)}$ and $H^*={\rm SO}^*(16)$. The group $H_c$
is ${\rm U}(8)={\rm U}(1)_E\times {\rm SU}(8)$ and ${\bf R}_c={\bf
28}_{+1}+\overline{{\bf 28}}_{-1}$, the subscripts being the ${\rm
U}(1)_E$ gradings. In this case we can choose complex basis elements
$K_M$, labeled by antisymmetric couples $[ab]$, $a,b=1,\dots, 8$:
$(K_M)=(K_{ab},\,K^{ab})$.

The Lax operator $L$ is defined to be an element of $\mathfrak{K}$.
Its components along the generators $K_M$ are the central and matter
charges of the four-dimensional theory, which depend on the
quantized charges and the scalar fields. For instance in the maximal
theory
\begin{eqnarray}
L\vert_{\mathfrak{K}_{{\bf R}_c}}&=&\sqrt{2}\,\left(\bar{Z}^{ab}\,K_{ab}+Z_{ab}\,K^{ab}\right)\,,
\end{eqnarray}
$Z_{ab}$ being the complex central charges of the four-dimensional
$\mathcal{N}=8$ parent theory. It is known that a complex $8\times
8$  matrix can be skew-diagonalized by means of a ${\rm U}(8)$
transformation. The $4$ real skew-eigenvalues $\rho_k$ of the
central charge matrix $Z_{ab}$ define the \emph{normal form} of the
four-dimensional central charges with respect to the action of
$H_c$. From a three-dimensional point of view, this procedure
amounts to rotating $L\vert_{\mathfrak{K}_{{\bf R}_c}}$, by means of
the adjoint action of $H_c$, into a $4$-dimensional abelian
subalgebra $\mathfrak{K}_N=\{K_k\}$ of $\mathfrak{K}_{{\bf R}_c}$
\begin{eqnarray}
\exists h\in H_c&:& h^{-1}\,L\vert_{\mathfrak{K}_{{\bf R}_c}}\,h=\sqrt{2}\,\sum_{k=1}^4\rho_k\,K_k\,.
\end{eqnarray}

We can generalize the above discussion to a generic theory and
define the normal form of  ${\bf R}_c$ with respect to $H_c$ as the
minimal subspace (\emph{normal subspace}) of its basis into which
any of its elements can be rotated by means of an
$H_c$-transformation. Let us denote by $p$ the dimension of this
subspace. Thus $p$ is the minimal number of charges into which the
most general set of central and matter charges, in ${\bf R}_c$, can
be reduced  by means of an $H_c$-transformation. In the maximal
theory we have seen that $p=4$. This implies that the most general
black hole solution modulo the action of the three-dimensional
global symmetry group $G$ (the \emph{generating} or \emph{seed
solution}) is a $p$-charge solution. Taking $\mathfrak{K}_{{\bf
R}_c}$ as the basis of the ${\bf R}_c$-representation, it was shown
in \cite{Bergshoeff:2008be} that its normal subspace
$\mathfrak{K}_N=\{K_\ell\}$ is spanned by $p$-commuting generators
$K_\ell$, $\ell=1,\dots, p$, where $p$ is the rank of the symmetric
Riemannian coset $H^*/H_c$: $p={\rm rank}(\frac{H^*}{H_c})$. In the
maximal theory, $p={\rm rank}(\frac{H^*}{H_c})={\rm rank}(\frac{{\rm
SO}^*(16)}{{\rm U}(8)})=4$. Similarly we can define a space
$\mathfrak{H}_{{\bf R}_c}$ of non-compact generators from the Cartan
decomposition of $\mathfrak{H}^*$ with respect to the lie algebra
$\mathfrak{H}_c$ of $H_c$: $\mathfrak{H}^*=\mathfrak{H}_c\oplus
\mathfrak{H}_{{\bf R}_c}$. Just as $\mathfrak{K}_{{\bf R}_c}$, also
$\mathfrak{H}_{{\bf R}_c}$ defines, with respect to the adjoint
action of $H_c$, the basis of a representation ${\bf R}_c$. We can
then define a normal subspace $\mathfrak{H}_N=\{J_\ell\}$,
$\ell=1,\dots, p$, of $\mathfrak{H}_{{\bf R}_c}$  with respect to
the action of $H_c$. The non-compact generators $J_\ell$ of the
normal subspace, can be chosen together with the compact
counterparts $K_\ell$ in $\mathfrak{K}_{{\bf R}_c}$ so that,
$K_\ell, \,J_\ell$ and $H_\ell\equiv [K_\ell,\,J_\ell]$ generate $p$
commuting $SL(2,\mathbb{R})$ subgroups of $G$. These groups define a
subspace of $G/H^*$ of the form
\begin{eqnarray}
\left(\frac{{\rm SL}(2,\mathbb{R})}{{\rm SO}(1,1)}\right)^p\subset \frac{G}{H^*}\,,
\end{eqnarray}
which consists of the $p$-fold product of $dS_2$ spaces. This space
and its generating isometries plays a role in the construction of
the seed solutions, relative to the action of $G$, of regular black
holes \cite{Bergshoeff:2008be}.

\subsection{The solvable parametrization and the Noether charges} \label{ssec:solvpar}
What we would like to stress from the above discussion is that the
$p$ commuting compact generators $K_\ell$ in the coset generate a
$p$-torus $T^p$ inside $G/H^*$. In the case of a $dS_2$ space, for
instance, $p=1$ and the torus is the non-trivial 1-cycle $S_1$ of
the corresponding hyperboloid. A generic element of $T^p$ can not in
general be brought to the form of a solvable group element of
$\exp{Solv}$ times an element of $H^*$. This can be easily seen in
the case of the space $dS_2$: The circle $S_1$ at fixed global time
interpolates between two solvable patches, each describing a copy of
Einstein's static universe (see \cite{Chemissany:2009hq} for a
detailed discussion of this issue). In general each 1-cycle of $T^p$
interpolates between different solvable patches in $G/H^*$. If one
solvable patch is described by a coset representative
$\mathbb{L}(\phi)\in \exp(Solv)$, the manifold being homogeneous, we
can find a compact element $\mathcal{D}$  of $G/H^*$ which maps the
origin of this patch into a point in any other solvable patch, this
being  described by a coset representative of the form
$\mathcal{D}\cdot\mathbb{L}(\phi) \in \mathcal{D}\cdot\exp(Solv)$.
Thus homogeneity of $G/H^*$ implies that any two solvable patches
are isometric images of each other. As pointed out in
\cite{Bossard:2009at}, the subset of points covered by the solvable
patches is dense in $G/H^*$ and the points which are at the boundary
of the various patches, and thus are not described by the solvable
parametrization, are those in which $\e^{-U}=0$, where $U$ is the
scalar entering the metric in the ansatz (\ref{reductionAnsatz}),
and that parametrizes the Cartan generator $T_0$ of $Solv$, see
below. Solutions crossing the boundary between two solvable patches
are characterized by the property that $\e^{-U}$ vanishes at some
finite $\tau$, and are therefore singular. It was shown in
\cite{Breitenlohner:1987dg} for geodesics in  $dS_2$, that only the
space-like ones cross the boundary of the solvable patch, while the
other geodesics unfold within the same solvable patch. Physical
fields belong to a single solvable patch (\emph{physical patch})
and, as long as we are interested in regular solutions only, we can
restrict ourselves to the physical patch only, since in doing so, we
are not loosing any physics. We can then define the scalar manifold
as a solvable symmetric pseudo-Riemannian manifold obtained by
identifying all solvable patches of $G/H^*$:
\begin{eqnarray}
{\mathcal{M}}_{scal}&\equiv & D\backslash G/H^*\,,
\end{eqnarray}
where $D$ is the discrete subgroup in $G/H^*$ whose elements map
solvable patches into one another.

Let us now define our solvable parametrization. It is convenient to
define the coset representative $\mathbb{L}$ in the solvable gauge
as follows
\begin{eqnarray}\label{reprbh}
\mathbb{L}(\phi^i)&=&\e^{-a\,E_{\beta_0}}\,\e^{\sqrt{2}\,Z^M\,T_M}\,\mathbb{L}_4(\phi^r)\,\e^{U\,H_0}\,,\label{Lgen}
\end{eqnarray}
where $\mathbb{L}_4$ is the  coset representative of $G_4/H_4$.
Using the coset representative (\ref{reprbh}) one can calculate the
sigma model metric, the Lax operator $L$ and Lax connection $W$
explicitly.

The structure constants of $Solv$, in the basis $\{T_0,T_r,T_M,T_\bullet\}$,
have the following non-vanishing components
\begin{eqnarray}
[T_A,\,T_B]&=&f_{AB}{}^C\,T_C\,;\nonumber\\
f_{0M}{}^N&=&\frac{1}{2}\,\delta_M^N\,\,;\,\,\,f_{0\bullet}{}^\bullet=1\,\,;\,\,\,f_{rs}{}^t
\,\,;\,\,\,f_{rM}{}^N=-(T_r)_M{}^N\,\,;\,\,\,f_{MN}{}^\bullet=\mathbb{C}_{MN}\,.
\end{eqnarray}
\par We find, after some calculations,
\begin{eqnarray}
\mathbb{L}_0{}^0&=&1\,\,;\,\,\,\mathbb{L}_0{}^M=\frac{\e^{-U}}{\sqrt{2}}\,Z^N\,\mathbb{L}_{4\,N}{}^M\,\,;\,\,\,\mathbb{L}_0{}^\bullet
=-\e^{-2\,U}a\,\,;\,\,\,\mathbb{L}_r{}^{{\hat{s}}}=\mathbb{L}_{4\,r}{}^{\hat{s}}\,,\nonumber\\\mathbb{L}_r{}^M&=&-\sqrt{2}\,\e^{-U}\,
({\bf
Z}^T\,T_r\,\mathbb{L}_4)^M\,\,;\,\,\,\mathbb{L}_r{}^\bullet=-\e^{-2\,U}\,T_{r\,MN}\,Z^M\,Z^N\,\,;\nonumber\\\mathbb{L}_M{}^N&=&
\e^{-U}\,\mathbb{L}_{4\,M}{}^N\,,\quad \mathbb{L}_M{}^\bullet
=\sqrt{2}\,\e^{-2\,U}\,\mathbb{C}_{MN}\,Z^N\,\,;\,\,\,
\mathbb{L}_\bullet{}^\bullet=\e^{-2\,U}\,.\label{l3}
\end{eqnarray}
In (\ref{l3}) we have used the $d=4$ coset representative
$\mathbb{L}_4$ and the $Solv_4$ generators $\{T_r\}$ in two different
representations: the $Solv_4$--adjoint representation
($\mathbb{L}_{4\,r}{}^{\hat{s}},\,(T_r)_{s}{}^t$) and the symplectic
representation of the electric-magnetic charges
($\mathbb{L}_{4\,M}{}^N,\,(T_r)_{M}{}^N$)
\begin{eqnarray}
\mathbb{L}_4^{-1}\,T_r\,\mathbb{L}_4&=&\mathbb{L}_{4\,r}{}^{\hat{s}}\,T_{\hat{s}}\,\,\,;\,\,\,
\mathbb{L}_4^{-1}\,T_M\,\mathbb{L}_4=\mathbb{L}_{4\,M}{}^N\,T_N\,,\nonumber\\
\left[T_r,\,T_s\right]&=&-(T_r)_{s}{}^t\,T_t\,\,\,;\,\,\,\left[T_r,\,T_M\right]=-(T_r)_{M}{}^N\,T_N\,\,\,;\,\,\,T_{r\,MN}\equiv
(T_r)_{M}{}^P\,\mathbb{C}_{PN}\,.
\end{eqnarray}
Recalling the definition of the Lax components  $Y^A$,
$\mathbb{L}^{-1}\,\dot{\mathbb{L}}=Y^A\,T_A$, we can now derive
their explicit form in terms of $\dot{\phi}^i$
\begin{eqnarray}\label{LAX}
Y^0&=&2\,\dot{U}\,\,;\,\,\,\,Y^{\hat{r}}=\sqrt{2} V_s{}^{\hat{r}}\,\dot{\phi}^s\,\,;\nonumber\\
Y^M&=&\sqrt{2}\,\e^{-U}\,\mathbb{L}_{4\,N}{}^M\,\dot{Z}^N\,\,;\nonumber\\
Y^\bullet&=&-\e^{-2\,U}\,(\dot{a}+{\bf Z}^T\,\mathbb{C}\,\dot{{\bf
Z}})\,,
\end{eqnarray}
where $V_r{}^{\hat{s}}\,d\phi^r$ is the vielbein of $G_4/H_4$ in
the solvable coordinates,
$G_{rs}=V_r{}^{\hat{t}}\,V_s{}^{\hat{u}}\,\delta_{\hat{t}\hat{u}}$.
From the Lax operator we compute the sigma model metric
\begin{equation}\label{sigma}
g_{ij}\,\dot{\phi}^i\,\dot{\phi}^j =
\alpha(Y^A\,Y^B\,\tilde{\eta}_{AB})\,.
\end{equation}
The normalisation coefficient $\alpha$ can be fixed by comparing
with the action obtained from dimensional reduction, as given by
equation (\ref{sigma3D}) in appendix \ref{APPENDIX}
\begin{equation}
\alpha= \frac{1}{2\tilde{\eta}_{00}}\,.
\end{equation}
We must furthermore have that
\begin{equation}\label{normalisations}
\tilde{\eta}_{NM}=-\tilde{\eta}_{00}\delta_{NM}\,,\qquad
\tilde{\eta}_{\bullet\bullet}=\tilde{\eta}_{00}\,,\qquad
\tilde{\eta}_{rs}=\tilde{\eta}_{00}\delta_{rs}\,,
\end{equation}
which can always be achieved by rescalings of the generators. If we
insist on the normalisations for the structure constants as given by
equation (\ref{algebra}) we generically loose the freedom to rescale
the generators such that $\tilde{\eta}_{00}=1$. For this reason we
explicitly carry factors of $\tilde{\eta}_{00}$ around in our
expressions.

We can now write the explicit form of $Q_A=\mathbb{L}_A{}^B\,Y_B$
\begin{eqnarray}
\tilde{\eta}_{00}^{-1}Q_0&=&2\,\dot{U} - \e^{-2\,U}\,{\bf
Z}^T\,\mathcal{M}_4\,\dot{{\bf
Z}}+ n\,a=2M_{ADM}\,,\nonumber\\
\tilde{\eta}_{00}^{-1}Q_r&=&\sqrt{2}(\mathbb{L}_4)_{r}{}^{\hat{s}}\,V_t{}^{\hat{u}}\dot{\phi}^t\delta_{\hat{s}\hat{u}}\,+2\,\e^{-2\,U}\,({\bf
Z}^T\,T_r\,\mathcal{M}_4\,\dot{{\bf
Z}})+T_{r\,MN}\,\,Z^M\,Z^N n\,,\nonumber\\
\tilde{\eta}_{00}^{-1}Q_M&=&-\sqrt{2}\,\left(\e^{-2\,U}\,\mathcal{M}_{4\,MN}\dot{
Z}^N+\mathbb{C}_{MN}\,Z^N n
\right)=\sqrt{2}\,\mathbb{C}_{MN}\,\mathcal{Q}^N\,,\nonumber\\
\tilde{\eta}_{00}^{-1}Q_\bullet&=&-n\,,\label{Noethercharges}
\end{eqnarray}
where $\mathcal{Q}=(m^I,\,e_I)$ is the vector of quantized charges
and, as usual we define $\mathcal{M}_{4\,MN}\equiv
\mathbb{L}_{4\,M}{}^P\,\mathbb{L}_{4\,N}{}^P>0$. It is useful to
express the Noether charges in terms of the conjugate momenta $P_i$:
$P_U,\,P_r,\,P_M,\,P_\bullet$. We know that
$P_i=g_{ij}\,\dot{\phi}^j$, which explicitly reads (see appendix)
\begin{eqnarray}
P_U&=&2\dot{U}\,\,,\,\,\,P_r=G_{rs}\,\dot{\phi}^s\,\,,\,\,\,P_M=-e^{-2\,U}\,\mathcal{M}_{4\,MN}\dot{Z}^N
-\tfrac{1}{2}n\,\mathbb{C}_{MN}Z^N\,\,,\,\,\,P_\bullet=\tfrac{1}{2}n\,.\nonumber
\end{eqnarray}
We find, after some algebra
\begin{eqnarray}
\tilde{\eta}_{00}^{-1}Q_0&=& P_U+ Z^M\,P_M + 2a\,P_\bullet\,\,,\nonumber\\
\tilde{\eta}_{00}^{-1}Q_r&=&\sqrt{2} (\mathbb{L}_4)_{r}^{\,\hat{s}}\,V^{-1}_{\hat{s}}{}^t\,P_t - 2P_M T_{rN}^{\,\,\,\,\,\,M}Z^N\,,\nonumber\\
\mathcal{Q}^M&=& \mathbb{C}^{MN}\,P_N - Z^MP_\bullet\,,\nonumber\\
\tilde{\eta}_{00}^{-1}Q_\bullet&=&-2P_\bullet\,.
\end{eqnarray}
\subsection{The (Fake) Superpotentials $\mathcal{W}_4,\,\mathcal{W}_3$}
The (fake) superpotential  $\mathcal{W}_4$ mentioned in the Introduction, is solution to the Hamilton-Jacobi 
equation associated to the autonomous system (\ref{effective1}) describing the radial flow of the four-dimensional scalar fields and warp factor:
 \begin{eqnarray}
\frac{1}{2}\,(\partial_U\mathcal{W}_4)^2+
\partial_r \mathcal{W}_4\partial^r \mathcal{W}_4&=& 2\,\mathrm{e}^{2\,U}\,V\,,
 \end{eqnarray}
 while $\mathcal{W}_3\equiv \mathcal{W}$ is solution to eq. (\ref{HJW}), associated with the three-dimensional Hamiltonian system.
The relation between the  superpotentials $\mathcal{W}_4,\,\mathcal{W}_3\equiv \mathcal{W}$ was found in \cite{Andrianopoli:2009je} to be:
\begin{eqnarray}
\mathcal{W}(\phi^i)&=&\mathcal{W}_4(U,\phi^r;\,m,e)+Z^M\,\mathbb{C}_{MN}\,\mathcal{Q}^N=\mathcal{W}_4+Z_I\,m^I-Z^I\,e_I\,.
\end{eqnarray}

\subsection{The Hamiltonians} Let us review, in brief, the algorithm introduced in
\cite{Fre:2009et} for constructing the Hamiltonians in involution associated with the Lax equation 
for the $\Sl(N,\mathbb{R})$--models\footnote{As anticipated in the Introduction, this algorithm
also works for the STU model.} (a more general construction will be given in \cite{MARIO.PIETRO.SASHA}).\par 
 In order to construct the
Hamiltonians, one makes use of the embedding (\ref{slembed}). Let us
introduce two indices
\begin{eqnarray}
a & = & 0, \cdots, \left[\frac{N}{2}\right] \,, \nonumber \\
\alpha & = & 1, \cdots, N - 2 a \,.
\end{eqnarray}
We will denote the Hamiltonians $\mathcal{H}(Y_A)$ by
$\mathfrak{h}_{a \alpha}$. The functions $\mathfrak{h}_{a \alpha}$
are then obtained by the relation
\begin{eqnarray}\label{hamdet}
& & \mathrm{det} \left\{ (L- \lambda \mathbbm{1})_{ij} : a+1 \leq i \leq N \,, 1 \leq j \leq N-a \right\} \nonumber \\
& & = \mathcal{E}_{a 0} \left( \lambda^{N-2a} +
\sum_{\alpha=1}^{N-2a} \mathfrak{h}_{a \alpha} \lambda^{N-2a -
\alpha}\right) \,, \quad a = 0, \cdots, \left[\frac{N}{2}\right] \,,
\end{eqnarray}
where $\mathcal{E}_{a0}$ is the coefficient of the power
$\lambda^{N-2a}$. One can check that the coefficient
$\mathfrak{h}_{01}$ corresponds to $\mathrm{Tr}(L)$ and is therefore
zero. The Hamiltonians $\mathfrak{h}_{0 \alpha}$ are essentially the
coefficients of the characteristic polynomial of $L$ and are
polynomial in the $Y_A$-variables. More specifically,
$\mathfrak{h}_{02}$ is quadratic and corresponds to the Hamiltonian function $\mathcal{H}$ defined in (\ref{defPH}). For $a \neq 0$, the
$\mathfrak{h}_{a \alpha}$ are rational functions of $Y_A$.

\subsection{Regularity}\label{reg} It is important to realise that
not every geodesic corresponds to a (physically reasonable) black
hole solution. We already mentioned that the solutions with $v^2<0$
are over-extreme. There is a convenient way to parametrise those $Q$
that correspond to regular black hole solutions\footnote{Irregular
solutions are not considered physical unless the naked singularity
is lightlike such that it coincides with the horizon (small black
holes). These cases are still of interest since one expects that
derivative corrections change the solution such as to develop a
horizon to cloak the singularity.}. This is based on a theorem due
to Breitenlohner, Gibbons and Maison \cite{Breitenlohner:1987dg}.
This theorem states that all regular non-extremal black hole
solutions can be transformed to the Schwarzschild solution using
$G$-transformations. This theorem can be rephrased on the level of
the matrix $Q$, as shown in \cite{Bossard:2009at}.  For the
Schwarschild solution, $U$ is the only scalar that is turned on.
Hence we have
$\mathbb{L}=\rme^{UH_0}$ and $\mathbbm{M}=\eta\rme^{2UH_0}$. 
Let us denote the Noether charge that belongs to the Schwarzschild
solution by $Q_s$, then we have
\begin{equation}
Q_s=2v H_0\,.
\end{equation}
In the adjoint representation we have the peculiar property that
$H_0$ has a 5 grading structure. This means that $H_0$ is
proportional to a diagonal matrix with only $-2, -1, 0, +1, +2$ as
entries. If we fix the normalisation of the generators such that
$H_0$ contains only the $-2, -1, 0, +1, +2$ as entries then
\begin{equation}
Q_s^5=20\,v^2\, Q_s^3 - 64\,v^4\,Q_s\,.
\end{equation}
Since $Q$ transforms in the adjoint of $G_3$: $Q\rightarrow
gQg^{-1}$, we find that the above relation is fulfilled for all
charge matrices that are connected with $Q_s$ and hence for all
regular non-extremal black hole solutions. For $\mathrm{E}_8$ we have the
particular property that the adjoint representation is also the
fundamental representation. But for all other groups encountered in
supergravity this is not the case and the particular property of
these groups is that the fundamental representation has a 3-grading
of $H_0$; this implies the simpler equation in the fundamental
\begin{equation}\label{regular1}
Q_s^3 =4v^2 Q_s\,.
\end{equation}

In sum, we have that regular black holes obey
\begin{equation}
Q^5=20\,v^2\, Q^3 - 64\,v^4\,Q\,.
\end{equation}
in the adjoint and
\begin{equation}\label{fund}
Q^3 =4v^2 Q\,.
\end{equation}
in the fundamental for all algebras of interest but $\mathrm{E}_8$. From this
we find the elegant result that extremal regular black holes are
described by nilpotent matrices for which the expansion $\e^{Q\tau}$
terminates after a finite number of steps leading to polynomial
expressions for the entries of $\mathbb{M}$ and hence to simpler
expressions for the scalars. It was conjectured in
\cite{Bossard:2009at} that in the limit $v\rightarrow 0$ another
regularity condition exists, which can be written in terms of
weighted Dynkin diagrams, but we do not investigate this further.

The regularity condition also facilitates the construction of
explicit solutions. If one constructs explicit solutions using the
formalism of the symmetric coset matrix $\mathbb{M}$, then
regularity implies
\begin{equation}
\mathbb{M}=\mathbb{M}(0)\Bigl(\mathbbm{1} +\tau Q
+\tfrac{1}{2}\tau^2 Q^2\Bigr)\,.
\end{equation}
when $v=0$ and
\begin{equation}
\mathbb{M}=\mathbb{M}(0)\Bigl(\mathbbm{1} +\frac{\sinh(2v\tau)}{2v}Q
+\frac{\cosh(2v\tau)-1}{4v^2} Q^2\Bigr)\,.
\end{equation}
when $v\neq 0$. Similarly in the Lax formalism of section 4, where
the computations rely on the matrix $ \mathcal{C}(\tau)\,=
\rme^{-2\, \tau \, L_0}$, which expands in the same way as the
matrix $\mathbb{M}$ given above.

On the level of the Hamiltonians the regularity of the solutions
becomes straightforward. First we observe that the polynomial
Hamiltonians are invariant under $G$. Hence, from Mazur's theorem,
we deduce that the polynomial Hamiltonians take the same value as
they do for the Schwarzschild solution. This means that a necessary
condition for regularity is that \emph{all polynomial Hamiltonians,
but the quadratic one ($\mathcal{H} \propto v^2$), are zero.}
This condition is also sufficient. For that we observe that the
polynomial Hamiltonians are the coefficients of the characteristic
polynomial $P_L(\lambda)$
\begin{equation}\label{caley}
P_L(\lambda) : \qquad  \lambda^N
+\mathcal{H}_2(L)\lambda^{N-2}+\mathcal{H}_3(L)\lambda^{N-3}+\ldots
+\mathcal{H}_{N}\,.
\end{equation}
If we use the Cayley--Hamilton theorem, $P_L(L)=0$, together with
the fact that all $\mathcal{H}_i=0$ with $i>2$, we find
\begin{equation}
L^N +\mathcal{H}_2L^{N-2}=0\,.
\end{equation}
We cannot multiply this equation with $L^{3-N}$ to find the
regularity condition (\ref{fund}) since $L$ is generically not
invertible (and especially not when it corresponds to a regular
solution). However the above equation does imply that the
eigenvalues $\lambda$ of $L$ obey
\begin{equation}
\lambda^n+\mathcal{H}_2\lambda^{n-2}=0\,,
\end{equation}
with solutions $\lambda=\pm\sqrt{\mathcal{H}_2}, 0$. This however
implies the 3-grading structure and hence the regularity condition
(\ref{fund})\footnote{The regularity condition (\ref{fund}) implies
that the solution is on the Schwarzschild orbit, although we have
not yet proven this in that direction. This is however
straightforward since the eigenvalues of $L$ have to  be $0,
\sqrt{\mathcal{H}_2}$ and $-\sqrt{\mathcal{H}_2}$. The
multiplicities of the eigenvalues is deduced using Tr$L=0$ and
Tr$L^2=2\mathcal{H}_2$. Indeed $0$ has multiplicity $N-2$ and the
other two eigenvalues each have multiplicity 1, showing that it is
the Schwarzschild solution.}.

\subsection{Normal forms} The above discussion is an example of how
normal forms are useful. By definition a normal form is the simplest
form of $Q\in \frak{g}\ominus\frak{H}$ obtained under the adjoint
action of $H^*$. Thus, the Schwarzschild solution is the normal form
of all regular non-extremal solutions. Another useful aspect of
normal forms comes from the fact that supersymmetry commutes with
$G$ and hence the study of the supersymmetry properties of the
normal form suffices to understand the supersymmetry of the general
solution.

In \cite{Bergshoeff:2008be} the normal form of $Q$ for a generic
geodesic was derived.  This is more general then the extended
theorem of Mazur \cite{Breitenlohner:1987dg} since it applies also
to extremal solutions and to singular solutions.

The theorem of \cite{Bergshoeff:2008be} states that the normal form
$Q_N$ is given by the following element of the algebra $\frak{g}$:
\begin{equation}
Q_N = \oplus_{i=1}^p[\frak{sl}(2)\ominus \frak{so}(1,1)]\quad
\oplus_{j=1}^q \frak{so}(1,1) \oplus Nil.
\end{equation}
The number $p$ is the rank of
the coset $H^*/H_c$, with $H_c$ the maximal compact subgroup of $H^*$.
\emph{Nil} is a nilpotent generator that commutes with the first
part of the normal form. In \cite{Bergshoeff:2008be} the focus was
on solutions where \emph{Nil} is absent and the generating solution
is described by $p$ axion-dilaton pairs and $q$ dilatons. The reason
is that these solutions are limits of non-extremal solutions with a
(complex) diagonalisable initial condition. It was also noted that
the known attractor solutions in $\mathcal{N}=8$ supergravity could
be understood from the normal form without \emph{Nil}. Regularity
and the absence of naked singularities implies that none of the $p$
axion-dilaton pairs can have negative velocity squared. Hence
extremal solutions are obtained by taking the $p$ axion-dilaton
pairs to be lightlike and to truncate the $q$ decoupled dilaton
scalars. The susy properties of the various solutions for the
$\mathcal{N}=8$ theory (and the related $\mathcal{N}=2$ STU model)
have been investigated using the normal form in
\cite{Bergshoeff:2008be}, see also \cite{Bossard:2009at,
Bossard:2009bw, Bossard:2009mz, Bossard:2010mv} for more recent
work.

\subsection{The dilatonic black hole}\label{slthree}
Consider the Einstein--Maxwell--dilaton action
\begin{equation}\label{EMD}
S=\int\sqrt{|g|}\Bigl(\tfrac{1}{2}\mathcal{R}-\tfrac{1}{2}(\partial\phi)^2
-\tfrac{1}{4}\rme^{a\phi}F^2 \Bigl)\,.
\end{equation}
This is probably the easiest theory that can describe black holes
with scalar hair. When the dilaton coupling $a$ obeys $a^2=6$ then
the action (\ref{EMD}) is just the circle reduction of pure gravity
in $d=5$ \footnote{In the language of the general sigma model
treated in the appendix we have $G_{rs}=1$,
$\mu_{IJ}=\e^{2\sqrt{3}\,\phi}$ and $\nu_{IJ}=0$.}. This model can
therefore be embedded in many supergravity theories (so does the
$a^2=2$ theory). The reduction over time leads to the
$\Sl(3,\Real)/\SO(2,1)$-coset. The black hole solutions of this
theory have been considered by many authors before
\cite{Dobiasch:1981vh, Gibbons:1982ih, Gaiotto:2007ag, Perz:2008kh}.
Here we redo the analysis in our framework because it nicely
illustrates our formalism.

\subsubsection{The $\frac{\Sl(3,\Real)}{\SO(2,1)}$ sigma
model}\label{SL3}

We define the coset element as in equation (\ref{reprbh}) and
consider the following fundamental representation for the generators
\begin{align}
&T_0=\frac{1}{2}\begin{pmatrix}1 & 0 &0 \\
0 & 0& 0 \\
0 & 0  & -1
 \end{pmatrix}\,,\qquad T_{\bullet}=\begin{pmatrix}0 & 0 & 1 \\
 0 & 0 & 0 \\
 0 & 0 & 0 \end{pmatrix}\,,\qquad T_{\phi}=\frac{1}{2\sqrt3}\begin{pmatrix} 1& 0 & 0 \\
 0 & -2 & 0 \\
 0 & 0 & 1 \end{pmatrix}\,,\nonumber\\
&T_1=\begin{pmatrix}0 & 0 & 0 \\
 0 & 0 & 1 \\
 0 & 0 & 0 \end{pmatrix}\,,\qquad\quad T_2=\begin{pmatrix}0 & 1 & 0 \\
 0 & 0 & 0 \\
 0 & 0 & 0 \end{pmatrix}\,.
\end{align}
The notation in the above expressions is as before, where $M$ runs
over $\{1,2\}$ and, since the solvable algebra in $d=4$ is trivial,
we have denoted $T_r=T_{\phi}$. From these solvable generators we
can construct the coset generators $K_A$ via
$K_A=\tfrac{1}{2}(T_A+\eta T_A^T\eta)$ with $\eta$ given by
\begin{equation}
\eta=\begin{pmatrix} 1 & 0 & 0\\
0 & -1 & 0\\
0 & 0  & 1 \end{pmatrix}\,.
\end{equation}
This allows us to compute the coset metric
$\tilde{\eta}_{AB}=\mathrm{Tr}(K_AK_B)$
\begin{equation}
\tilde{\eta}=\text{diag}(\tilde{\eta}_{00}, \tilde{\eta}_{\phi\phi},
\tilde{\eta}_{11}, \tilde{\eta}_{22},
\tilde{\eta}_{\bullet\bullet})=\text{diag}(\tfrac{1}{2},
\tfrac{1}{2}, -\tfrac{1}{2}, -\tfrac{1}{2}, -\tfrac{1}{2})\,.
\end{equation}
This coset metric indeed satisfies the normalisations
(\ref{normalisations}). The coset representative (\ref{reprbh}) is
explicitly given by
$\mathbb{L}=\e^{-aT_{\bullet}}\e^{\sqrt{2}Z^MT_M}\e^{2\sqrt{2}\phi
T_{\phi}}\e^{2UT_0}$. In components this is
\begin{equation}
\mathbb{L}=
\begin{pmatrix}
\e^{U + \tfrac{\sqrt2}{\sqrt{3}}\phi} &
\sqrt{2}\e^{\tfrac{-2\sqrt2}{\sqrt{3}}\phi} Z^2 & \e^{-U +
\tfrac{\sqrt2}{\sqrt{3}}\phi } (-a + Z^1 Z^2)\\
0 & \e^{\tfrac{-2\sqrt2}{\sqrt{3}}\phi} & \sqrt{2}\e^{-U +
\tfrac{\sqrt2}{\sqrt{3}}\phi } Z^1\\
0 & 0 & \e^{-U + \tfrac{\sqrt2}{\sqrt{3}}\phi }\\
\end{pmatrix}\,.
\end{equation}
From this we can compute the Lax operator
\begin{equation}L=
\begin{pmatrix}
\tfrac{1}{2}Y^0 +\tfrac{1}{2\sqrt{3}} Y^{\phi} & \tfrac{1}{2}Y^2 &
\tfrac{1}{2}Y^{\bullet}\\
-\tfrac{1}{2}Y^2 & -\tfrac{1}{\sqrt{3}} Y^{\phi} & \tfrac{1}{2}Y^1\\
\tfrac{1}{2}Y^{\bullet} & -\tfrac{1}{2}Y^1 & -\tfrac{1}{2}Y^0
+\tfrac{1}{2\sqrt{3}} Y^{\phi}
\end{pmatrix}\,,
\end{equation}
with
\begin{align}
& Y^0 = 2 \dot{U}\,,\qquad  Y^{\phi}=\sqrt2\dot{\phi}\,,\nonumber\\
& Y^1=\sqrt{2}\,\e^{-U + \sqrt{6}\phi}\dot{Z}^1\,,\nonumber\\
& Y^2=\sqrt{2}\,\e^{-U - \sqrt{6}\phi}\dot{Z}^2\,,\nonumber\\
&Y^{\bullet}=- \e^{-2U} (\dot{a} + Z^2\dot{Z}^1 - Z^1 \dot{Z}^2)\,.
\end{align}
\subsubsection{The solutions}

In order to investigate regular solutions of this model, we first
show how the regularity condition (\ref{regular1}) can be
simplified. As the initial Lax operator $L_0 = L(\tau = 0)$ is an
element of the $\Sl(3)$ algebra, it obeys the following identity
\begin{equation}
L_0^3 = \frac{1}{2} \mathrm{Tr}(L_0^2) L_0 + \frac{1}{3}
\mathrm{Tr}(L_0^3) \mathbbm{1} \,.
\end{equation}
From (\ref{regular1}), one thus sees that the condition for having
regular solutions is simply
\begin{equation} \label{regsl3}
\mathrm{Tr}(L_0^3) = 0 \,.
\end{equation}
Of course, this is the $\Sl(3)$-version of the statement that all
polynomial Hamiltonians, but the quadratic one, have to vanish for
regular solutions (see the discussion around equation
(\ref{caley})). Practically this means that the dilaton charge can
be written in terms of the mass and the electric-magnetic charges,
something that was known before \cite{Gibbons:1982ih}. In the
extremal case one has
\begin{equation}\label{extrem}
\mathrm{Tr}(L_0^2)=0\qquad\Rightarrow\qquad \upsilon=0\,,
\end{equation}
we then immediately find that $L_0$ is nilpotent of degree 3 :
$L_0^3=0$. This allows only 2 arbitrary charges since
$\mathrm{Tr}(L_0^3)=\mathrm{Tr}(L_0^2)=0$ give 2 conditions for 4
charges.

In order to find solutions using the integration algorithms, we will
use a specific parametrization for the initial condition $L_0$,
inspired from \cite{Gibbons:1982ih}, that solves the regularity
condition
\begin{equation}
L_{0}=\label{ParaL0}\left(%
\begin{array}{ccc}  \frac{\alpha }{2}&\frac{-1}{2}\sqrt{\frac{{\alpha }^3- 4v^2\,\alpha }{\alpha  - \beta
}}&0\\ \frac{1}{2}\sqrt{\frac{{\alpha }^3-4v^2\,\alpha }{\alpha -
\beta }}&\frac{-1}{2}(\alpha  +\beta)
  & \frac{-1}{2}\sqrt{\frac{{\beta }^3- 4v^2\,\beta   }{\beta-\alpha }}\\0&\frac{1}{2}
  \sqrt{\frac{{\beta }^3- 4v^2\,\beta
}{\beta-\alpha  }}&\frac{\beta }{2}
\end{array}%
\right)\,.
\end{equation}
We can choose the gauge in which $\mathbb{L}_3 (0)=\mathbbm{1}$.
This implies that $U(0)=1$, which is the usual coordinate choice,
$Z^M(0)=a(0)=0$, which can be consistently obtained, when $n=0$, by
shifts of the axions. Only the choice $\phi(0)=0$ is a restriction.
However, if we use the dilatation symmetry in 4 dimensions:
\begin{equation}
\phi\rightarrow \phi+c\,,\qquad F_{\tau t} \rightarrow F_{\tau
t}\rme^{\sqrt{6}c}\,,\qquad F_{\theta\phi} \rightarrow F_{\theta
\phi}\rme^{-\sqrt{6}c} \,.
\end{equation}
we can generate the solutions with $\phi(0)\neq 0$.

Let us solve first for the non-extremal solutions. The initial Lax
operator is given by (\ref{ParaL0}). We present the solution for the
initial condition $\phi(0)=0.$  In terms of the two functions
$B(\tau),\,C(\tau)$
\begin{eqnarray}\label{STfunctions} B(\tau)&=&\frac{\alpha \,\left( 4v^2 - {\beta }^2 \right)  + \beta \,\left( -4v^2 + \alpha \,\beta  \right) \,\cosh (2v\,\tau) +
    2v\,\left( \alpha  - \beta  \right) \,\beta \,\sinh (2v\,\tau)}{4v^2\,\left( \alpha  - \beta  \right) }\,,\nonumber\\
       C(\tau)&=&\frac{\left( -4v^2 + {\alpha }^2 \right) \,\beta  + \alpha \,\left( 4v^2 - \alpha \,\beta  \right) \,\cosh (2v\,\tau) +
    2v\,\alpha \,\left( -\alpha  + \beta  \right) \,\sinh (2v\,\tau)}{4v^2\,\left( \alpha  - \beta  \right)
    }\,, \nonumber \\ \end{eqnarray}
the solution reads
\begin{equation}\label{solution}
\e^{2U}= (BC)^{-1/2},\qquad
\e^{\phi}=(B^{-1}C)^{-\frac{\sqrt{3}}{2\sqrt2}}\,.
\end{equation}

In the extremal limit, defined by $\upsilon=0$, the corresponding
solution is given by equation (\ref{solution}), where $B$ and $C$
now read
\begin{equation}
B(\tau)=1+\beta\tau+\frac{\alpha\beta^{2}}{2(\alpha-\beta)}\tau^{2}\,, \qquad
C(\tau)=1-\alpha\tau-\frac{\alpha^{2}\beta}{2(\alpha-\beta)}\tau^{2}.
\end{equation}
We can take the limit where either the electric or the magnetic
charge vanishes. Then the solution simplifies considerably and the
regular horizon collapses to coincide with the black hole
singularity: the solution becomes a small black hole. One can
readily check that the nilpotency degree of $L_0$ becomes 2 instead
of 3 in this limit.

With $\mathbb{L}_3(0)=\mathbbm{1}$ we have $Q = L_0$. Through
equations (\ref{Noethercharges}), this allows us to give a physical
interpretation to $\alpha$ and $\beta$
\begin{align}
& Q_0= \tfrac{1}{4}(\alpha-\beta)=M\,,\\
& Q_{\bullet}=0=-\tfrac{1}{2}n\,,\\
& Q_1=\tfrac{1}{2}\sqrt{\frac{\beta^3-4v^2\beta}{\beta-\alpha}}=-\tfrac{1}{\sqrt{2}}\mathcal{Q}_{electric}\,,\\
& Q_2=\tfrac{1}{2}\sqrt{\frac{\alpha^3-4v^2\alpha}{\alpha-\beta}}=\tfrac{1}{\sqrt{2}}\mathcal{Q}_{magnetic}\,,\\
& Q_{\phi}=\tfrac{\sqrt{3}}{2}(\alpha+\beta)\,.
\end{align}

\subsubsection{The normal form}
Since the scalar manifold in three dimensions is
$\Sl(3,\Real)/\SO(2,1)$, the normal form of $L_0$ should be given by
\begin{equation}
L_0\in[\frak{sl}(2)\ominus \frak{so}(1,1)]\oplus\frak{so}(1,1)\oplus
Nil\,.
\end{equation}
For extremal solutions the decoupled $\frak{so}(1,1)$ should be
zero. One verifies that the $\frac{\frak{sl}(2)}{\frak{so}(2,1)}$
part has nilpotency degree 2 in the fundamental. This implies that
we need the extra nilpotent generator \emph{Nil} of degree 3 in
order to have a regular solution. Indeed, the solution without
\emph{Nil} corresponds to the small black hole solution that has
either vanishing magnetic or electric charge. Furthermore, if one
also truncates the lightlike charges in the
$\frac{\frak{sl}(2)}{\frak{so}(2,1)}$ part, such that one is left
with the \emph{Nil} generator, only then one recovers the
\emph{double extremal} solution. This is the solution for which the
dilaton is constant everywhere since its initial value is at the
minimum of the black hole potential and its initial velocity is
zero. This occurs when $\alpha=-\beta$.

\subsubsection{Hamiltonian viewpoint}
Following the procedure outlined in section \ref{integrability}, we
can construct the Hamiltonians in involution needed for
integrability, two of which depend on the $Y$ \footnote{Note that we
are restricting ourselves here to the case where the Taub-NUT charge
vanishes. In case one allows for a non-zero Taub-NUT charge, the
procedure of section 3 leads to an extra Hamiltonian, as would be
required by integrability in that case. This extra Hamiltonian is a
rational function of the $Y^A$ and corresponds to a Casimir, in the
terminology of section 3.3. For vanishing Taub-NUT charge, this
extra Hamiltonian is however not present.}
\begin{eqnarray} \label{hampart1}
\mathcal{H}_{1}(Y)&=&
-\tfrac{1}{4}(Y^0)^2 + \tfrac{1}{4}(Y^1)^2 + \tfrac{1}{4}(Y^2)^2 - \tfrac{1}{4} (Y^{\phi})^2 \\
\mathcal{H}_{2}(Y)&=& \tfrac{1}{8}Y^0\bigl((Y^2)^2-(Y^1)^2\bigr)
-\tfrac{1}{8\sqrt3}Y^{\phi}\bigl((Y^1)^2 + (Y^2)^2\bigr)\nonumber\\
&&+\tfrac{1}{12\sqrt{3}}(Y^{\phi})^3-\tfrac{1}{4\sqrt{3}}(Y^0)^2Y^{\phi}\,.
\end{eqnarray}
The first of these, $\mathcal{H}_1(Y^A)$ corresponds to $\mathcal{H}
= -\frac{1}{2} Y_A Y^A$ and vanishes for extremal solutions. The
second Hamiltonian, $\mathcal{H}_2(Y^A)$, is cubic and proportional
to $\mathrm{Tr}(L_0^3)$. It vanishes for regular solutions.

The proof of integrability requires two extra Hamiltonians, that now
depend on the charges $Q_A$. In principle they are obtained by
making the substitution $Y^A \rightarrow Q_A$ in
$\mathcal{H}_1(Y^A)$, $\mathcal{H}_2(Y^A)$. This however leads to
rather complicated expressions. Since we are considering vanishing
Taub-NUT charge here, we can adopt a more practical approach. We
simply choose the two electro-magnetic charges as Hamiltonians
\begin{equation} \label{hampart2}
\mathcal{H}_{3}(Q_A)=Q_{1}\,,\qquad \mathcal{H}_{4}(Q_A)=Q_{2}.
\end{equation}
These still Poisson-commute with the $Y^A$ variables and hence with
$\mathcal{H}_1$ and $\mathcal{H}_2$. Moreover, for vanishing
Taub-NUT charge, they also Poisson-commute among themselves since
the electric and magnetic charges close a Heisenberg algebra with
the Taub-NUT charge. For vanishing Taub-NUT charge this Heisenberg
algebra degenerates and $\mathcal{H}_3$ and $\mathcal{H}_4$ commute.
On-shell, the Hamiltonians take on constant values
$\mathcal{H}_i(t)\equiv h_i$. The interpretations of these
Hamiltonians is now straightforward. The first Hamiltonian $h_1$
corresponds to the extremality parameter. The second Hamiltonian
keeps track of the regularity condition; if $h_2=0$ ($h_2\neq 0$)
the solution is regular (singular). The third and fourth hamiltonian
are proportional to the magnetic and electric charge. Hence, regular
extremal solutions are parameterized by two numbers $(Q_{electric},
Q_{magnetic})$ and regular non-extremal solutions are parameterized
by three numbers $(Q_{electric}, Q_{magnetic})$ and $h_1$.

Recalling the sigma-model
\begin{equation}
\mathcal{L}=\tfrac{1}{2}g_{ij}\dot{\phi}^i\dot{\phi}^j= \dot{U}^2 +
\tfrac{1}{2}\dot{\phi}^2
-\tfrac{1}{2}\e^{-2U+\sqrt{6}\phi}(\dot{Z}^1)^2 -
\tfrac{1}{2}\e^{-2U-\sqrt{6}\phi}(\dot{Z}^2)^2 \,,
\end{equation}
where the generalised coordinates are $U,\phi, Z^1, Z^2$, the
conjugate momenta are given by
\begin{equation}
P_U=2\dot{U}\,,\qquad P_{\phi}=\dot{\phi}\,,\qquad
P_1=-\e^{-2U+\sqrt{6}\phi}\dot{Z}^1\,,\qquad P_2=-
\e^{-2U-\sqrt{6}\phi}\dot{Z}^2\,.
\end{equation}
The four non-vanishing Hamiltonians in (\ref{hampart1}) and
(\ref{hampart2}) may then be expressed in terms of the coordinates
and momenta as follows
\begin{eqnarray}
\mathcal{H}_1&=&-\tfrac{1}{4}P_U^2 -\tfrac{1}{2}P_{\phi}^2 + \tfrac{1}{2}\e^{2U-\sqrt{6}\phi}P_1^2 +\tfrac{1}{2}\e^{2U+\sqrt{6}\phi}P_2^2 \nonumber \,,\\
\mathcal{H}_2&=&-\tfrac{1}{2\sqrt{6}}P_U^2P_{\phi}  +   \tfrac{1}{3\sqrt{6}}P_{\phi}^3  -  \tfrac{1}{4}\e^{2U-\sqrt6\phi}P_UP_1^2 - \tfrac{1}{2\sqrt{6}}\e^{2U-\sqrt6\phi}P_{\phi}P_1^2 \nonumber\\
&&+ \tfrac{1}{4}\e^{2U+\sqrt6\phi}P_UP_2^2 -\tfrac{1}{2\sqrt{6}}\e^{2U+\sqrt6\phi}P_{\phi}P_2^2  \nonumber \,,\\
\mathcal{H}_3&=& \tfrac{\sqrt{2}}{2}P_1\,,\qquad \mathcal{H}_4 = \tfrac{\sqrt{2}}{2}P_2\,.
\end{eqnarray}
We can write the momenta $P_A$ in terms of the coordinates  by
solving the 4 equations
\begin{equation}
h_i=\mathcal{H}_i(P_A, \{U,\phi, Z^1, Z^2\})\,,
\end{equation}
in terms of the $P_A$
\begin{equation}\label{firstorder}
P_A=f_A(U,\phi, Z^1, Z^2)\,,
\end{equation}
when the Jacobian $J=\textrm{det}\left(\frac{\partial
\mathcal{H}_{i}}{\partial P_{j}}\right)$ is non-zero. For the
example at hand we find
\begin{eqnarray} J&=&
-\tfrac{1}{8} e^{2 U- \sqrt{6} \phi}P_{1}^2 P_{\phi}+\tfrac{1}{8\sqrt{6}} e^{2
U- \sqrt{6} \phi} P_{1}^2 P_{U}+\tfrac{1}{8} e^{ \sqrt{6} \phi+2 U} P_{2}^2
P_{\phi} +\nonumber\\ &&\tfrac{1}{8\sqrt{6}} e^{ \sqrt{6} \phi+2 U}
P_{2}^2P_{U} -\tfrac{ \sqrt{6}}{8}P_{\phi}^2
P_{U}+\tfrac{1}{8\sqrt{6}}P_{U}^3\,.\end{eqnarray} From this explicit
expression one can infer that the regularity condition $h_{2}=0$ can
\emph{not} be satisfied when $J=0.$ Hence, for physical solutions
$h_2=0$, it makes sense to define the first order equation
(\ref{firstorder}).

It then turns out that solving explicitly for the $P_A$ in terms of
the coordinates is the easiest for the extremal solutions $h_1=0$.
The result is
\begin{eqnarray}
P_U&=&\tfrac{1}{2}\e^{\sqrt{\tfrac{3}{2}}\phi+U}\Bigl(h_4^{3/2} +\e^{-\tfrac{4}{\sqrt{6}}\phi}h_3^{2/3} \Bigr)^{3/2} \,,\\
P_{\phi}&=&\tfrac{\sqrt{6}}{4}\e^U\Bigl(\e^{\sqrt{\tfrac{3}{2}}\phi}h_4^{2/3}-\e^{-\tfrac{1}{\sqrt6}\phi}h_3^{2/3}
\Bigr)\sqrt{h_4^{2/3} + \e^{-\tfrac{4}{\sqrt6}\phi}h_3^{2/3}}\,.
\end{eqnarray}
We can explicitly integrate this to find the 4$d$ fake
superpotential
\begin{equation}\label{Wextremal}
\mathcal{W}_4(U,\phi)\equiv 2\, \rme^U\,W(\phi)\,,\qquad
W(\phi)=\frac{1}{2}\,\Bigl(\e^{\tfrac{2}{\sqrt6}\phi}h_4^{2/3} +
\e^{-\tfrac{2}{\sqrt6}\phi}h_3^{2/3} \Bigr)^{3/2}\,.
\end{equation}
As expected the fake superpotential factorizes with the black hole
warp factor $U$ \cite{Perz:2008kh}.

In the non-extremal case the expressions for $P_A$ in terms of the
coordinates are page filling unless we take either the magnetic or
electric charge equal to zero. When the electric charge is zero we
find
\begin{eqnarray}
P_U&=& -\tfrac{1}{2}\sqrt{\e^{\sqrt6\phi+2U}h_4^2-16h_1}\,,\\
P_{\phi}&=&-\tfrac{\sqrt6}{4}\sqrt{\e^{\sqrt6\phi+2U}h_4^2-16h_1}\,,
\end{eqnarray}
where $h_1<0$. The 4$d$ fake superpotential then becomes
\begin{align}
\mathcal{W}_4(U,\phi)=&\sqrt{\e^{\sqrt6\phi+2U}h_4^2-16h_1}
+\sqrt{-h_1}\log\Bigl(-16h_1+4\sqrt{-h_1}\sqrt{\e^{\sqrt6\phi+2U}h_4^2-16h_1}
\Bigr)\nonumber\\ &-4\sqrt{-h_1}(U+\sqrt{\tfrac{3}{2}}\phi) \,.
\end{align}
This indeed coincides with the expressions found in
\cite{Perz:2008kh} and in the limit $h_1\rightarrow 0$ it collapses
to the single charge expression for the extremal superpotential
(\ref{Wextremal})
\begin{equation}
\mathcal{W}_4(U,\phi)=\e^U\e^{\sqrt{\tfrac{3}{2}}\phi}h_4 \,.
\end{equation}

\section{Discussion}

Let us summarize the main results obtained in this paper. We have
established new insights in the solvability and integrability of the
geodesic equations of motion on symmetric coset spaces that appear
as sigma models of supergravity theories that are reduced over the
timelike direction.

Concerning the solvability we have presented a recursive but closed
formula for the coset representative describing a generic geodesic solution to
equation (\ref{solutioncosetrepr}). This extends the
existing results that only supply a general formula for the Lax
operator \cite{Fre:2003ep, Chemissany:2009hq, Fre:2009et,
Chemissany:2009af, kodama-1995,
kodama2-1995,kodama3-1995,Fre:2009dg}. Finding the coset
representative still implied solving first order differential
equations in that approach. It is not always fully appreciated in
the literature that the solvability was not yet established for the
coset representative $\mathbb{L}$. This is due to the fact that the
solution for the symmetric combination $\mathbb{M}=\mathbb{L}\eta
\mathbb{L}^T$ can be obtained from exponentiating a Lie algebra
element. However, in order to extract $\mathbb{L}$ in the solvable
parametrisation, which defines the physical fields \cite{Andrianopoli:1996bq}, from this symmetric combination requires solving an involved
 equation for the compensator. This problem is circumvented by the 
 algorithm we propose.

Concerning the issue of formal integrability we have proven in a constructive way 
that the second order equations  of motion, i.e. the full autonomous Hamiltonian system, 
 are integrable in the sense of Liouville, once the same property has been established 
  for the first order Lax pair problem.
This means that we have proven the existence of $n$ constants of
motion that mutually Poisson-commute, where $n$ is the dimension of
the symmetric space, starting from those associated with the Lax pair problem. 
This extends the previous result which had
proven Liouville integrability of the first order differential
equations obeyed by the tangent to the geodesic. These results solve
an open-standing question about the existence of a fake
superpotential for black hole solutions. The fake superpotential
description is the supergravity-way of describing Hamilton--Jacobi
integrability and since Liouville integrability implies the latter
we have proven the (local) existence of a fake superpotential for all
stationary, spherically symmetric, black hole solutions to symmetric
supergravity theories. The obstruction to a global existence of $\mathcal{W}$
is due to the existence of points in which the Jacobian 
det$(\partial h_i/\partial P_j)\neq 0$ vanishes. We have proven in an 
explicit example that this Jacobian is always non-vanishing in the subspace 
spanned by regular solutions, thus ensuring  therein a global existence of
 $\mathcal{W}$. We expect this to be the case for generic symmetric 
 geometries, though a general proof is missing.

We also presented an explicit construction for the $n$ constants of
motion leaning on earlier obtained results \cite{Fre:2009et}.
The constants are divided in constants that depend solely on $Y,$
$\mathcal{H}_{\alpha}(Y),$ and constants that depend solely on
 $Q,$ $\mathcal{H}_{a}(Q),$ where $Y$ are
 the Lax components and the $Q$ are the Noether charge components in the
 solvable directions. The constants that depend on $Y$ can be divided in
 two sets : $\mathcal{H}_{\alpha}(Y) = \{\mathcal{H}_{\ell}(Y), \mathcal{H}_{a}(Y)\},$
  where the $\mathcal{H}_{\ell}(Y)$ correspond
 to Casimirs. Among the $\mathcal{H}_{\alpha}(Y),$ there are a certain number of
polynomial constants. These have a simple interpretation. One of them
corresponds to the geodesic velocity, which is proportional to the
 extremality parameter ST. The remaining polynomial $\mathcal{H}_{\alpha}(Y)$ then
 keep track of the regularity of the solution: they have to vanish
 for a regular solution. Among the remaining $\mathcal{H}_{\alpha}(Y),$ there
 is one constant that corresponds to the Taub-NUT charge. So far we have
 been unable to give a physical interpretation to the remaining
 constants in the set $\mathcal{H}_{\alpha}(Y).$  The constants $\mathcal{H}_{a}(Q)$
 are functions of  the electro-magnetic charges $(e^{I},m_{I})$ and have no
 easy interpretation. However,  for most cases of interest the Taub-NUT charge
 vanishes and then we have  shown that one can replace $\mathcal{H}_{a}(Q)$ for the pure
 electro-magnetic charges, making the physical interpretation of
  all constants, but  the remaining ones in $\mathcal{H}_{\alpha}(Y),$ clear.  We have explicitly
worked this out for the dilatonic black holes that arise in
Kaluza-Klein theories since this is the easiest possible example to
demonstrate our findings with. This example has only one
non-polynomial $\mathcal{H}(Y ),$ which is the Taub-NUT charge, so that we were
unable to learn about the physical meaning of other possible
non-polynomial Hamiltonians from this example. We
anticipate to investigate the Hamiltonians  in
more involved models, such as the STU model, which should make the
physical interpretation clear.

Finally our analysis on Liouville integrability can have a bearing 
on the quantum description of black holes along the lines of
 \cite{Bossard:2009we,Gunaydin:2007bg}. In these papers
 the authors developed a quantum description of black hole
 based on the quantization of the radial evolution of the scalar fields and the 
 warp factor. The resulting wave function was suggested to have, in the 
 semi-classical limit, the following form:
 \begin{eqnarray}
 \Psi(\phi^i)&\propto & \exp\left(i\,\mathcal{W}(\phi^i)\right)=\exp\left(i\,\mathcal{W}_4(U, \phi^r)+i\,(Z_I\,m^I-Z^I\,e_I)\right)\,.
 \end{eqnarray}
In this picture the $n$ Hamiltonians in involutions, which we have been dealing with in the present paper, 
 will provide a complete set of commuting observables, in terms of which to completely characterize the black hole state.\par  Understanding the physical meaning of these Hamiltonians is therefore an important step 
 in order to understand the physics of black holes in supergravity.

\section*{Acknowledgements}
W.C. is supported in part by the Natural Sciences and Engineering
Research Council (NSERC) of Canada. The work of M.T. is supported by
the government grant PRIN 2007. T.V.R. is supported by the G\"oran
Gustafsson Foundation. J.R. and T.V.R. also like to thank the
Politecnico di Torino for its hospitality. The work of A.S. was
partially supported by the RFBR Grants No. $09-02-12417-ofi \_m$,
$09-02-00725-a$, $09-02-91349-NNIO\_a$; DFG grant No 436
$RUS/113/669$, and the Heisenberg-Landau Program.

\appendix

\section{The sigma model in $d=3+0$}\label{APPENDIX}
In this section we closely follow the seminal paper
\cite{Breitenlohner:1987dg}.

Consider the action (\ref{4Daction}). The equations of motion for
the vectors $B^I$ are
\begin{equation}
\d \bigl(\mu_{IJ}\star\d B^J -\nu_{IJ}\d B^J\bigr)=0\,,
\end{equation}
such that, locally, we can introduce the dual potentials $C_I$
\begin{equation}
\mu_{IJ}\star\d B^J -\nu_{IJ}\d B^J=\d C_I\,.
\end{equation}
The set of field strengths $(\d B, \d C)$ are not independent and
obey the following duality relation
\begin{equation}
\begin{pmatrix}
\d B\\ \d C
\end{pmatrix}= \mathbb{C} \mathcal{M}_4\star \begin{pmatrix}
\d B\\  \d C
\end{pmatrix}\,,
\end{equation}
where
\begin{equation}
\mathbb{C}=\begin{pmatrix}0 & -\mathbbm{1}\\
+\mathbbm{1} & 0 \end{pmatrix}\,,\qquad \mathcal{M}_4 =
\begin{pmatrix} \mu
+\nu\mu^{-1}\nu & \nu\mu^{-1}\\
\mu^{-1}\nu & \mu^{-1}
\end{pmatrix}\,.
\end{equation}
Let us now reduce the action over the timelike direction using the
ansatz (\ref{reductionAnsatz}). The reduced action then becomes
\begin{align}
S_3&=\int\Bigl(\tfrac{1}{2}\star R_3 - \star\d U\wedge\d U
+\tfrac{1}{4}\e^{4U}\star F_{KK}\wedge F_{KK} -
\tfrac{1}{2}G_{rs}\star\d \phi^r\wedge\d\phi^s \nonumber \\ & +
\tfrac{1}{2}\mu_{IJ}\e^{-2U}\star \d Z^I\wedge \d Z^J
-\tfrac{1}{2}\e^{2U}\mu_{IJ}\star(\tilde{G}^I+Z^IF_{KK})\wedge
(\tilde{G}^J+Z^JF_{KK})\nonumber
\\ & -\nu_{IJ}(\tilde{G}^I+Z^IF_{KK})\wedge \d Z^J\Bigr)\,,
\end{align}
The vectors $A_{KK}$ and $\tilde{B}^I$ can be dualised to scalars
$\chi$ and $Z_I$ by adding them as Lagrange multipliers to the
action that ensure the Bianchi identities
\begin{equation}
S'_3=S_3 +\chi \d F_{KK} + Z_I\d \tilde{G}^I\,.
\end{equation}
Varying the action $S_3'$ with respect to $F_{KK}$ and $\tilde{G}^I$
gives the equations of motion
\begin{align}
\d Z_J  &=-\e^{2U}\star \mu_{IJ}(\tilde{G}^I+Z^IF_{KK})-\nu_{IJ}\d Z^I \,,\label{dual1}\\
\d \chi &= \tfrac{1}{2}\e^{4U}\star F_{KK} +Z^I\d
Z_I\,.\label{dual2}
\end{align}
One can verify that the scalars $Z_I$ coincide with the timelike
components of the dual potentials $C_I$
\begin{equation}
C_I=\tilde{C}_I + Z_I(\d t+A_{KK})\,,
\end{equation}
where $\tilde{C}_I$ is a vector in $d=3$.

Dualisation of the action $S_3$ is equivalent to eliminating
$F_{KK}$ and $\tilde{G}^I$ from  the action $S_3'$ using the two
identities (\ref{dual1}, \ref{dual2}). If we furthermore make the
field redefinition
\begin{equation}
2\chi=a+Z^IZ_I\,,
\end{equation}
(such that $a$ is a symplectic invariant scalar) we find the
action\footnote{The sigma model metric corresponds to a Lorentzian
version of the metric obtained in \cite{Fre:2008zd}, where the
similar problem was considered for time-dependent solutions, i.e.
Euclidean sigma models. Once explicit solutions for the scalar
fields have been obtained, one can uplift these to four dimensions
using similar uplifting formula as developed in \cite{Fre:2008zd}.}
\begin{align}
S_3&=\int\Bigl(\tfrac{1}{2}\star R_3 - \star\d U\wedge\d U
-\tfrac{1}{2}G_{rs}\star\d \phi^r\wedge\d\phi^s +\tfrac{1}{2}\e^{-2U}\star \d {\bf Z}^T \wedge \mathcal{M}_4 \d {\bf Z} \nonumber\\
& -\tfrac{1}{4}\e^{-4U}\star(\d a + {\bf Z}^T \mathbb{C}\d {\bf
Z})\wedge (\d a + {\bf Z}^T \mathbb{C}\d {\bf
Z})\Bigr)\,.\label{sigma3D}
\end{align}
In the main text we have used the notation
\begin{equation}
{\bf Z}\equiv (Z^M)=(Z^I, Z_I)\,.
\end{equation}

In a compact notation we write
\begin{equation}
S_3=\int \tfrac{1}{2}\star R_3 -\tfrac{1}{2}
g_{ij}(\phi)\star\d\phi^i\wedge\d\phi^j\,.
\end{equation}

\section{Conventions for charges and mass}\label{APPENDIXB}
As usual we define charges via Gauss' law, for which we integrate a
closed 3-form over the spatial dimensions:
\begin{align}
\mathcal{Q}^I=m^I&=\tfrac{1}{4\pi} \int_3\d G^I \,\,\,= \tfrac{1}{4\pi}\int_{S^2}G^I\,,\\
\mathcal{Q}_I=e_I&=\tfrac{1}{4\pi} \int_3\d F_I \,\,\,\,= \tfrac{1}{4\pi}\int_{S_2}F_I\,,\\
 n&=\tfrac{1}{4\pi} \int_3\d F_{KK} = \tfrac{1}{4\pi} \int_{S_2}F_{KK}\,,
\end{align}
where
\begin{equation}
F_I=\d C_I=\mu_{IJ}\star G^J -\nu_{IJ}G^J\,.
\end{equation}
We have denoted magnetic charges  by $m^I$, electric charges by
$e_I$ and the Taub-NUT charge by $n$. There is a subtlety in the
definition of the electric and magnetic charge in a spacetime for
which $n\neq 0$. In that case the integrals that define the
$\mathcal{Q}$'s are dependent on the radial coordinate and hence not
constants\footnote{The killing one-form $e_t = \d t +A_KK$ is not
hyper surface orthogonal since it is not closed $\d e_t=F_{KK}\neq
0$. Hence we cannot define a spatial hypersurface orthogonal to the
Killing vector. }. The definition that gives constant charges, when
$n\neq 0$ can, for instance, be derived using Noethers theorem and
is given in the main text. From the formulas of appendix
\ref{APPENDIX} we find
\begin{align}
\mathcal{Q}^N&=\e^{-2U}(\mathbb{C}M\dot{\bold{Z}})^N\,,\\
 n&=\e^{-4U}(\dot{a}+\bold{Z}^T\mathbb{C}\dot{\bold{Z}})\,,
\end{align}
where we organised the electric and magnetic charges in the
symplectic vector
\begin{equation}
\mathcal{Q}^N=(m^I\,,\,\, e_I)\,.
\end{equation}

The mass of the black hole spacetime is defined by the Komar
integral
\begin{equation}
M=-\frac{1}{8\pi}\int_{S^2_{\infty}}\nabla^{\alpha}\xi^\beta\d
S_{\alpha\beta}\,,\qquad \d
S_{\alpha\beta}=-2n_{[\alpha}r_{\beta]}\sqrt{\sigma}\d\theta\d\phi\,,
\end{equation}
where $\xi^{\alpha}=\xi^t=1$ is the timelike killing vector,
$n^{\alpha}=n^t=\e^{-U}$ is the timelike normal to $S^2_{\infty}$,
$r^{\beta}=r^{\tau}=\e^{U-2A}$ is the spacelike (radial) normal to
$S^2_{\infty}$, and finally, $\sqrt{\sigma}=\sin\theta\e^{2A-2U}$ is
the determinant of the induced metric on $S^2_{\infty}$. We then
find
\begin{equation}
M=\Gamma^t_{t\,\tau}(0)=\dot{U}(0)\,.
\end{equation}

\bibliography{bhint22}
\bibliographystyle{utphysmodb}

\end{document}